\documentclass[11pt]{article}

\usepackage[table]{xcolor}
\usepackage[final]{acl}
\usepackage{times}
\usepackage{latexsym}
\usepackage[inkscapelatex=false]{svg}
\usepackage{tabularray}
\usepackage{subfigure}
\usepackage[T1]{fontenc}

\usepackage[utf8]{inputenc}

\usepackage{microtype}

\usepackage{inconsolata}

\usepackage{graphicx}

\usepackage{times}
\usepackage{latexsym}

\usepackage[T1]{fontenc}

\usepackage[utf8]{inputenc}

\usepackage{microtype}

\usepackage{inconsolata}

\usepackage{graphicx}

\usepackage{makecell}
\usepackage{bbding}
\usepackage{multirow}
\usepackage{pgfplots}
\usepackage{svg}
\usepackage{amsmath} 
\usepackage{amssymb}
\usepackage{url}
\usepackage{tablefootnote}
\usepackage{graphicx}
\usepackage{cuted}
\usepackage{tabularx}
\usepackage{xcolor}
\usepackage{pifont}
\usepackage{CJKutf8}
\usepackage{amssymb} 
\usepackage{cancel} 
\usepackage{hyperref}
\usepackage{adjustbox}

\definecolor{myblue}{HTML}{4E84C4}
\definecolor{myred}{HTML}{B02418}
\definecolor{mygreen}{HTML}{96C8A2}

\usepackage{listings}
\definecolor{deepgreen}{RGB}{6,153,6} 
\definecolor{deepred}{RGB}{254,34,35}    
\definecolor{deepyellow}{HTML}{FFBF00}

\title{TransBench: Breaking Barriers for Transferable Graphical User Interface Agents in Dynamic Digital Environments}

\author{
	Yuheng Lu\textsuperscript{1}\thanks{\quad Equal contribution}, Qian Yu\textsuperscript{1}\footnotemark[1], Hongru Wang\textsuperscript{2}\footnotemark[1], Zeming Liu\textsuperscript{1}\thanks{\quad Corresponding author: Zeming Liu.}, Wei Su\textsuperscript{3}, 
    \\ \textbf{
    Yanping Liu\textsuperscript{3}, Yuhang Guo\textsuperscript{3}, Maocheng Liang\textsuperscript{1}, Yunhong Wang\textsuperscript{1}, Haifeng Wang\textsuperscript{4}
    }\\
    	\textsuperscript{1}Beihang University, Beijing, China \hspace{1mm} \textsuperscript{2}The Chinese University of Hong Kong, China \hspace{1mm} \\
	\textsuperscript{3}Beijing Institute of Technology, Beijing, China \hspace{1mm}
	\textsuperscript{4}Baidu Inc., Beijing, China \hspace{1mm}\\
	{\tt \{20375061, qiantraceyyu, zmliu\}@buaa.edu.cn} 
	{\tt  hrwang@se.cuhk.edu.hk} \\
    }

\begin{document}
\maketitle

\begin{abstract}
Graphical User Interface (GUI) agents, which autonomously operate on digital interfaces through natural language instructions, hold transformative potential for accessibility, automation, and user experience. A critical aspect of their functionality is \textit{grounding} --- the ability to map linguistic intents to visual and structural interface elements. However, existing GUI agents often struggle to adapt to the dynamic and interconnected nature of real-world digital environments, where tasks frequently span multiple platforms and applications while also being impacted by version updates. To address this, we introduce \texttt{TransBench}, the first benchmark designed to systematically evaluate and enhance the transferability of GUI agents across three key dimensions: \textit{cross-version transferability} (adapting to version updates), \textit{cross-platform transferability} (generalizing across platforms like iOS, Android, and Web), and \textit{cross-application transferability} (handling tasks spanning functionally distinct apps). \texttt{TransBench} includes 15 app categories with diverse functionalities, capturing essential pages across versions and platforms to enable robust evaluation. Our experiments demonstrate significant improvements in grounding accuracy, showcasing the practical utility of GUI agents in dynamic, real-world environments. Our code and data will be publicly available at \href{https://github.com/BUAA-IRIP-LLM/TransBench}{\texttt{TransBench}}.
\end{abstract}

\begin{figure*}[!ht]
\vspace{-10pt}
    \centering
    \includegraphics[width=1\linewidth]{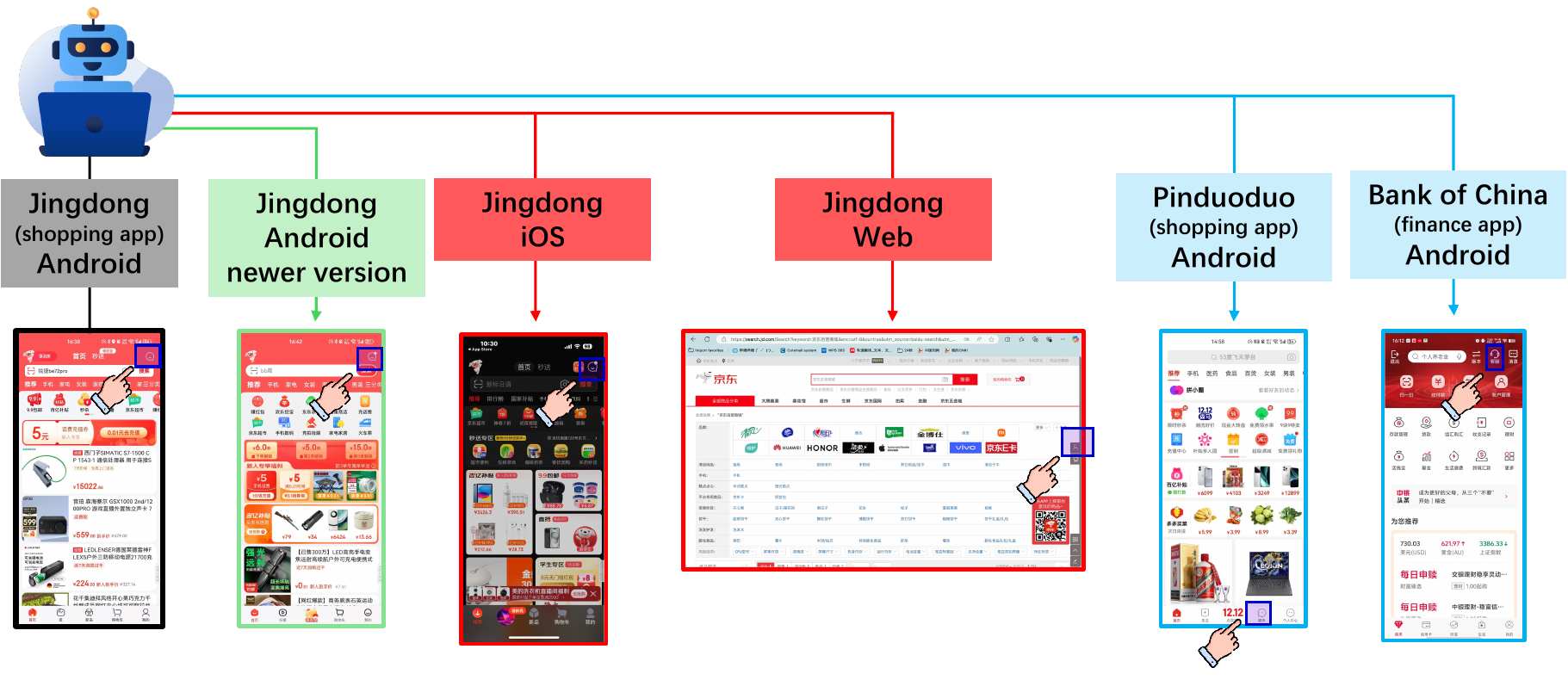}
    \caption{Interpretation of Transferability's three aspects. \textcolor{mygreen}{Green} means \textit{cross-version transferability}: transferring the knowledge learned from the homepage of Jingdong (a Chinese shopping app) from Android version 12.0.0 to a newer Android version, 13.6.8. \textcolor{myred}{Red} means \textit{cross-platform transferability}: transferring from the Android version of Jingdong to its iOS version 13.8.1 and Web version. \textcolor{myblue}{Blue} means \textit{cross-application transferability}: transferring from Jingdong to other apps with the same functionality (e.g., shopping: Pinduoduo) or with different functionality (e.g., Finance: Bank of China)}

    \label{fig:intro}
\end{figure*}

\section{Introduction}
GUI (Graphical User Interface) agents \cite{Zhang2024LargeLM}, which are autonomous agents acting in the digital world via operating on GUIs, enables users to accomplish complex tasks through natural language instructions \cite{Chen2024GUIWORLDAD,ma-etal-2024-coco,ijcai2024p339,hong2024cogagent,wu-etal-2024-mobilevlm, tool_tut}. These agents locate and manipulate multimodal GUI elements (i.e., buttons, icons, and menus) \cite{kapoor2025omniact} and autonomously execute corresponding operations (e.g., clicking, scrolling) \cite{gao2024assistgui} across diverse interfaces given the user instructions \cite{Lu2024OmniParserFP, mukhtar2025artificial}. By translating natural language instructions into precise actions, GUI agents democratize access to digital systems, offering transformative potential for accessibility, automation, and user experience. A core aspect of their functionality is \textit{grounding}: the ability to map linguistic intents from instruction to visual and structural interface components. It is crucial to ensure effective grounding, as failures in accurately interpreting and localizing GUI elements propagate to downstream execution errors, rendering even sophisticated action planning futile.

While prior work has advanced GUI agents' grounding capabilities, existing approaches focus exclusively on platform-specific settings, such as mobile apps (e.g., GUI-Odyssey \cite{lu2024gui}, AUITestAgent \cite{Hu2024AUITestAgentAR}), desktop interface (e.g., AssistGUI \cite{gao2024assistgui}), and web environments (e.g., Mind2web \cite{deng2024mind2web}). However, real-world applications operate dynamically: they span multiple platforms (e.g., iOS, Android, Web) and evolve continuously, with version updates frequently altering GUI layouts and functionalities. Meanwhile, user instructions often span applications with relevant or distinct functionalities, such as requesting "comparing products on \textit{Amazon} and \textit{Alibaba} with its review videos on \textit{Youtube}" and implicitly assume cross-version or cross-platform consistency. This challenge highlights a crucial aspect of grounding for GUI agents --- transferability. Without it, rigid version-, platform-, or app-specific grounding fails to generalize, making agents brittle in practice.

To address this issue, we first formally identify three levels of transferability in terms of grounding capability of GUI agents, as shown in Figure  \ref{fig:intro}: 1) \textit{cross-version transferability}: localizing GUI elements despite interface changes from version updates; 2) \textit{cross-platform transferability}: transferring grounding knowledge between platforms with divergent interaction patterns; and 3) \textit{cross-application transferability}: generalizing from interchangeable features to partially overlapping or functionally distinct ones. Therefore, these dimensions collectively determine whether the GUI agents can generalize beyond narrow, static settings to cross-version, cross-platform, and cross-application workflows.

To systematically evaluate and enhance the transferability at these three levels, we introduce \texttt{TransBench}, the first benchmark that addresses version and platform discrepancies with abundant real-world applications. Specifically, we carefully design a data collection pipeline in three consecutive steps: 1) screenshot collections, ranging from Android (including multiple versions), iOS \footnote{Old versions for iOS applications are unavailable.}, and web platforms; and 2) bounding boxes annotations; and 3) user instruction generation. We conduct rigorous quality control and human verification to ensure the quality and diversity of our benchmarks. We hope our dataset not only provides a robust foundation but also sets a new standard for future research in transferability, enabling the development of more adaptive and generalizable GUI agents across dynamic digital environments. Overall, our contributions can be summarized as follows:

\vspace{-6pt}
 \begin{itemize}
     \item We are the first to identify and highlight the challenge of transferability in grounding tasks, focusing on enabling GUI agents to execute multi-app tasks while adapting to version updates and platform differences.
     \item To support this, we design \texttt{TransBench}, a benchmark featuring datasets with screenshots of each application’s essential pages across versions and platforms, covering 15 categories with diverse functionalities.
     \item Using \texttt{TransBench}, our experiments showcase significant progress in grounding accuracy, highlighting increased robustness and practicality for transferability in dynamic digital environments.
 \end{itemize}

\section{Related work}
\subsection{GUI Agents Datasets}

GUI agent datasets play a crucial role in evaluating model performance and enhancing agents' abilities to understand GUI elements and execute tasks across diverse applications \cite{liu2023agentbench, chen2024guicourse, Liu2024VisualAgentBenchTL}. Most of the existing benchmarks tend to specialize in specific platforms, such as web \cite{deng2024mind2web}, mobile \cite{Hu2024AUITestAgentAR, lu2024gui}, and desktop \cite{gao2024assistgui}. For example, Mind2Web \cite{deng2024mind2web} is designed for web-based environments, while Android-specific datasets include PixelHelp \cite{li2020mapping}, MoTIF \cite{hwang2025motif}, GUI Odyssey \cite{lu2024gui}, and MobileViews \cite{gao2024mobileviews}. Only in recent works, such as VisualAgentBench \cite{liu2024visualagentbench} and WebHybrid \cite{gou2024navigating}, has the importance of cross-platform evaluation been widely recognized. However, they primarily concentrate on structural differences between different platforms, overlooking other levels of transferability for GUI agents, such as cross-version transferability and cross-application transferability. To the best of our knowledge, no existing dataset simultaneously tackles all these challenges. It is believed that our proposed \texttt{TransBench} can fulfill this blank by providing a benchmark that comprehensively and systematically evaluates GUI agents across these crucial dimensions.

\vspace{-.07pt}
\subsection{GUI Agents Models}

GUI agents play a crucial role in enabling intelligent automation, assisting users in navigating digital environments, and improving human-computer interaction \cite{Lu2024OmniParserFP, mukhtar2025artificial}. Recent advancements in GUI agent models have significantly improved their ability to understand complex interface layouts and user interactions \cite{wang2024gui}. For example, UI-BERT \cite{Bai2021UIBertLG} enhances agents' comprehension of user intent and interface structures by leveraging contextual representations, and AutoGLM \cite{liu2024autoglm} builds on this by integrating both textual and visual data, improving adaptability but at the cost of increased computational demands However, they still struggle with generalizing to unseen or frequently changing layouts. Besides that, MobileVLM \cite{chu2023mobilevlm} enhances task execution efficiency in mobile applications, while MobileAgent \cite{wang2024mobile} leverages multimodal data to handle multi-step commands. Furthermore, AutoMobileGPT \cite{yang2023auto} advances natural language task execution by enabling seamless interaction across diverse applications. As the need for GUI agents that can seamlessly adapt across different app versions, platforms, and functionalities continues to grow, advancing their transferability remains a crucial research challenge. Furthermore, addressing the complexities of varying user interfaces and interaction patterns will be key to ensuring robust performance in real-world applications.

\begin{figure*}[htbp]
    \centering
    \includegraphics[width=.9\linewidth]{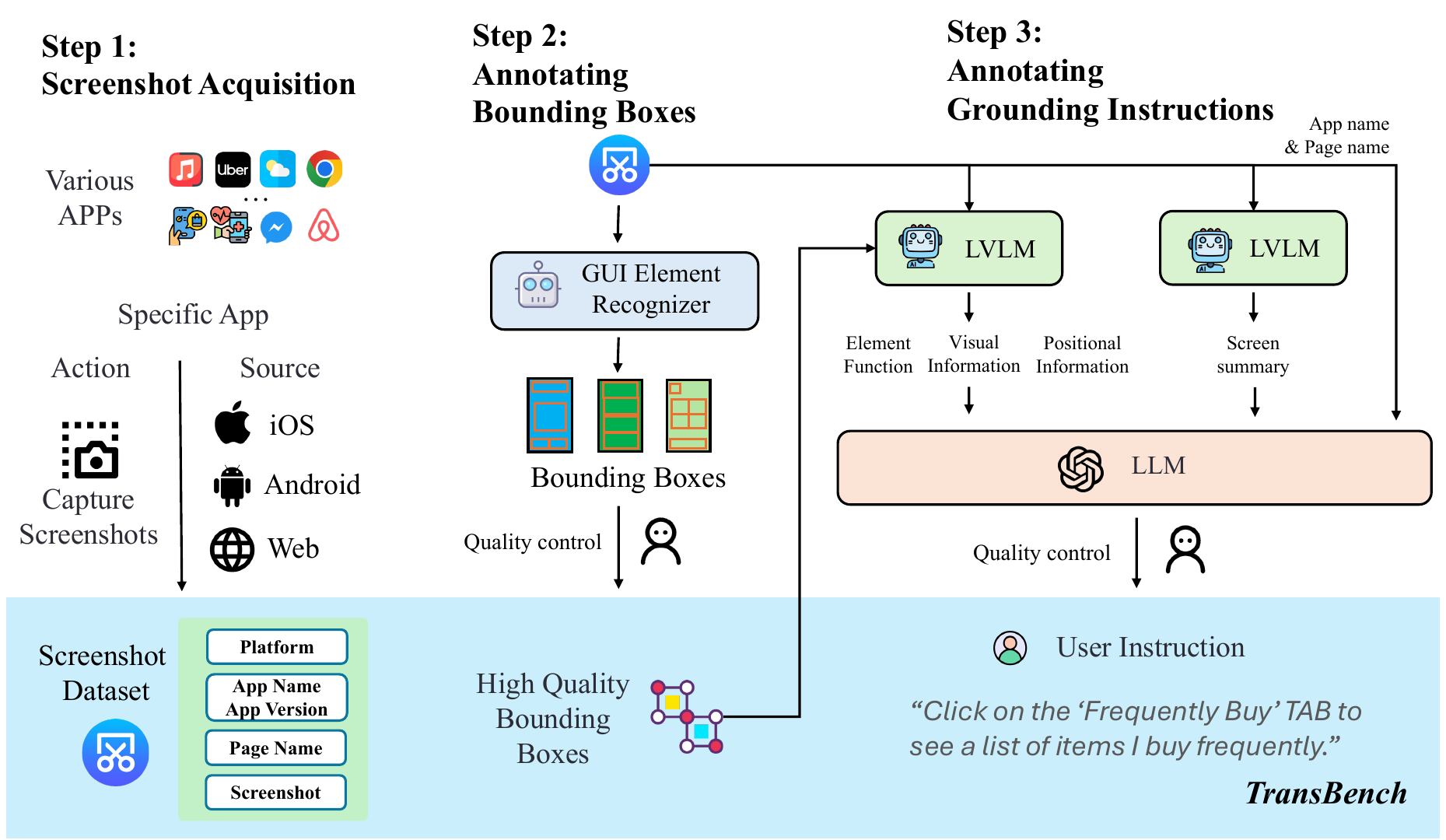}
    \caption{Interpretation of data collection process. The blue box represents our proposed benchmark -\texttt{TransBench}, which consists of three parts: ScreenShot Acquisition, Annotating Bounding Boxes, and Annotating Grounding Instructions. Platform means iOS, Android, and Web. Page names are manually divided into page names according to human semantics, such as "Shopping cart," "My page," "Home," "Comments," and so on, which usually have similar functions.}
    \label{fig:pipeline}
\vspace{-3mm}
\end{figure*}

\section{\texttt{TransBench} Construction}

It is difficult to directly leverage existing benchmarks as our seed dataset, since there are two significant challenges. First of all, current datasets are typically constructed as task sets and lack critical metadata, such as app names, page titles, version numbers, and platform details. However, these details are crucial for the study of transferability, as it depends on accurately identifying and comparing interface variations across different versions, platforms, and apps. Secondly, a comprehensive evaluation of transferability requires establishing correspondence relationships across multiple dimensions, such as mapping an app's current version to its previous versions and linking different platform-specific versions of the same app with the same pages. Existing datasets typically confine tasks within a single version or platform, making it infeasible to be reused as an evaluation of transferability. Therefore, we provide a detailed data collection pipeline for \texttt{TransBench} in this section (as shown in Figure \ref{fig:pipeline}), alongside the formal task definition. Table \ref{related_work} shows the detailed comparison between \texttt{TransBench} with other popular benchmarks.

\begin{table}[t]
\vspace{5mm}
\centering
    \tabcolsep=0.15cm
     \begin{adjustbox}{max width=0.47 \textwidth}
    \begin{tabular}{l|ccccc|c} 
    \hline

      \multirow{2}*{\textbf{Name}}&\multicolumn{5}{c|}{\textbf{Transferability}}&\multirow{2}*{\textbf{Lan}}\\
      
		\cline{2-6}
      ~& \multicolumn{3}{c}{\textit{Version}}& \textit{Platform}& \textit{Application}& ~\\
\hline
      
      Mind2Web \citep{deng2024mind2web}&\multicolumn{3}{c}{\color{deepred}\XSolidBrush}&\color{deepred}\XSolidBrush &\color{deepred}\XSolidBrush &en\\
    
    PixelHelp \citep{li2020mapping}& \multicolumn{3}{c}{\color{deepred}\XSolidBrush}& \color{deepred}\XSolidBrush & \color{deepred}\XSolidBrush &en\\
    
    MoTIF \citep{hwang2025motif}& 
    \multicolumn{3}{c}{\color{deepred}\XSolidBrush}& \color{deepred}\XSolidBrush & \color{deepred}\XSolidBrush &en\\
 
    GUI Odyssey \citep{lu2024gui}& \multicolumn{3}{c}{\color{deepred}\XSolidBrush}& \color{deepred}\XSolidBrush & \color{deepgreen}\CheckmarkBold&en\\
 
    E-ANT \citep{wang2024ant}& \multicolumn{3}{c}{\color{deepred}\XSolidBrush}& \color{deepred}\XSolidBrush & \color{deepred}\XSolidBrush &en\\
 
    Mobile3M \citep{wu2024mobilevlm}& \multicolumn{3}{c}{\color{deepred}\XSolidBrush}& \color{deepred}\XSolidBrush & \color{deepred}\XSolidBrush&ch\\
      
    MobileViews \citep{gao2024mobileviews}& \multicolumn{3}{c}{\color{deepred}\XSolidBrush}&\color{deepred}\XSolidBrush &\color{deepred}\XSolidBrush&ch\\
      
    VisualAgentBench \citep{liu2024visualagentbench}& \multicolumn{3}{c}{\color{deepred}\XSolidBrush}&\color{deepgreen}\CheckmarkBold&\color{deepred}\XSolidBrush &en\\
 
    WebHybrid \citep{gou2024navigating}& \multicolumn{3}{c}{\color{deepred}\XSolidBrush}& \color{deepgreen}\CheckmarkBold& \color{deepred}\XSolidBrush &en\\
      
    \hline
      
    \textbf{\textsc{TransBench}} (Ours)& \multicolumn{3}{c}{\color{deepgreen}\CheckmarkBold}& \color{deepgreen}\CheckmarkBold& \color{deepgreen}\CheckmarkBold&ch\\
\hline
\end{tabular}
\end{adjustbox}
 \caption{\label{related_work}
Comparison between TransBench to other GUI agent datasets from transferabilities' three aspects, including cross-version, cross-platform, and cross-application. "Lan" stands for "Language" and "ch" means targeting Chinese apps).
}
\label{tab:related_work}
\vspace{-2mm}
\end{table}

\subsection{Task Definition}
The input is a user instruction $u$ and a screenshot $s_i^j$ where $i$ stands for $i_{th}$ APP and $j$ means $j_{th}$ screenshot of this APP. Screenshot $s_i^j$ is combined with a set of instructions $\{u_1, u_2, ..., u_m\}$, which of each is associated with a ground truth bounding box $b_k = (x_{\min}, y_{\min}, x_{\max}, y_{\max})$. The agent's goal is to output a coordinate $(x, y)$. The prediction is correct if $(x, y)$ falls within the corresponding bounding box $b_k$. The objective is to improve the accuracy of GUI grounding across varying versions, platforms, and applications.

\subsection{Step 1: Screenshot Acquisition}

To cover diverse user instructions and applications, we identify 81 commonly used multi-platform applications in practice drawn from both previous studies and everyday usage, including shopping, video streaming, social networking, travel, lifestyle, maps, music, communication, finance, email, reading, education, camera, fitness, and utility tools. A complete list of applications is provided in the Appendix \ref{tab:app_name_desc}. Furthermore, considering the different levels of transferabilities, we define two types of screenshots: 1) fundamental screens that are common across most applications (i.e., homepage, message page, user profile page), allowing agents to establish a basic understanding of frequently encountered GUI elements; and 2) domain-specific screenshots which capture the unique functionalities of each application, enabling agents to adapt to specialized tasks. Consequently, we successfully collect the seed dataset, which includes a total of 1,459 screenshots: 825 from Android (covering both new and old versions), 429 from iOS, and 205 from web platforms. Tables listing the names of the applications and the titles of their corresponding pages are provided in the Appendix \ref{appendix1_name_title}.

\subsection{Step 2: Annotating Bounding Boxes}
Bounding boxes are essential for identifying and localizing GUI interface components such as buttons or input fields \cite{gou2024navigating}. In this way, we first utilize an automated annotation tool to generate preliminary bounding boxes, providing a foundational layer for the following work. After that, manual verification is conducted to address any discrepancies identified in the automated process.
\vspace{-3pt}
\paragraph{Automatic Annotations.} We use OmniParser \cite{Lu2024OmniParserFP} to automatically identify the bounding boxes due to its strong GUI element recognition capabilities. This automated process enables efficient identification of key GUI elements, reducing the manual workload in the initial stages. Additionally, an automated filtering process is applied to exclude non-essential elements, such as status bars or advertisements. This filtering step ensures that the dataset remains focused on GUI elements relevant for evaluating interaction capabilities.
\vspace{-3pt}
\paragraph{Manual Verification.} Following automated annotation, manual verification is performed by four well-educated human annotators. To streamline this process, we develop a specialized annotation tool, \textbf{GUILabeller}\footnote{Details of our tool can be found in Appendix \ref{appendix1_tool_detail}}, designed to facilitate flexible and efficient manual adjustments. Each human annotator is required to review each bounding box to identify whether it correctly encapsulates the intended GUI element. If discrepancies are found, the annotators manually adjust or redraw the bounding box to ensure precision. Additionally, particular attention is given to verifying the semantic equivalence of GUI components \cite{gou2024navigating}, which refers to cases where multiple GUI elements might belong to a larger GUI component and trigger identical outcomes. To address this issue, an additional larger bounding box is added during verification to encapsulate semantically equivalent elements, forming a hierarchical GUI element structure.
    

\subsection{Step 3: Annotating Grounding Instructions}
Following the completion of bounding box annotation, resulting in over 65,000 bounding boxes, the subsequent step is to generate high-quality grounding instructions while minimizing human intervention to ensure diversity and accuracy. This process is structured into three key steps, including extracting bounding box attributes, generating screen summaries, and constructing grounding instructions. Each step's prompt details can be found in Appendix~\ref{appendix1_prompt_details}\footnote{All prompts can be found in the Appendix if not stated.}. Manual verification is performed at the end to ensure correctness. Data examples can be found in Appendix  \ref{appendix:example}.

\paragraph{Bounding Box Attributes Acquisition.} Screenshots and their corresponding bounding boxes are processed to Qwen2VL \cite{Wang2024Qwen2VLEV} to obtain three-dimensional attributes (inspired by ARIA-UI \cite{yang2024aria}), including visual features, positional relationships, and functional characteristics.


\paragraph{Screen Summaries Generation.} Each screenshot, alongside its relevant metadata such as the application name and page title, is incorporated into prompts designed for Qwen2VL to generate screen summaries. These summaries synthesize both visual and contextual information, providing a holistic understanding of the interface's layout and functionality, which serves as the foundation for generating grounding instructions.

\paragraph{User Instruction Construction.} Using the bounding box attributes and screen summaries obtained from previous steps, we prompt Qwen-plus (for its strong reasoning, imagination, and instruction-following abilities) to construct the required user instruction. The generation process leverages multidimensional visual information provided by the visual model, as well as commonsense information associated with application names and page titles as prior knowledge. 

\paragraph{Quality Control.} Finally, to ensure data quality, manual verification is performed. Four human annotators together address inconsistencies and refine instructions while necessary. As a result, more than 22,000 high-quality grounding instructions are refined, and only these refined instructions are used in the following experiments.  
Furthermore, following \citet{liu-etal-2020-towards-conversational}, \citet{shi-etal-2023-midmed} and \citet{wang_survey_2024}, we employ human evaluations to assess data accuracy. An instruction is considered correct only if it precisely corresponds to the single corresponding bounding box, while pointing to more than one bounding box is considered incorrect. The final evaluation yields an average score of \textit{95.5\%}, confirming the high quality of the grounding instructions in the dataset.

\begin{table}[t]
\setlength{\belowcaptionskip}{0pt}
    \centering
    \begin{adjustbox}{max width=0.49\textwidth}

   \begin{tabular}{l |cccc }
    \hline
    \textbf{Statistics} & \textbf{Android old}& \textbf{Android new}& \textbf{iOS}& \textbf{Web}\\
    \hline
    \textbf{\# Apps}& 77& 80& 81& 47\\
     \hline
 \# Screenshots& 393& 432& 429& 205\\
 \# Bounding Boxes& 17,455& 19,384& 14,477&14,341\\
 \# Checked Instructions& 5,696& 6,305& 6,046&4,191\\
 \# Fundamental Pages& 300& 300& 300& 150\\
 \# Domain-Specific Pages& 93& 132& 129& 55\\
    \hline
    Avg. \# Screenshots& 5.1& 5.4& 5.3& 4.4\\
 Avg. \# Bounding Boxes& 226.7& 242.3& 178.7&305.1\\
 Avg. \# Instructions& 74.0& 78.8& 74.6&89.2\\
 Avg. \#Fundamental Pages& 3.9& 3.8& 3.7& 3.2\\
 Avg. \# Domain-Specific Pages& 1.2& 1.7& 1.6& 1.2\\
    \hline
    \end{tabular}
    \end{adjustbox}

    \caption{The data statistics of our proposed \texttt{TransBench}. Avg. meas average on single App. Bounding boxes include boxes with unchecked instructions and boxes with checked Instructions.}
    \label{tab:data_stat}
    \vspace{-4mm}
\end{table}

\subsection{Data Statistics}

Table \ref{tab:data_stat} illustrates the statistics of \texttt{TransBench}. Specifically, it includes up to 81 apps across Android (old and new versions), iOS, and Web platforms, with a total of 1,459 screenshots and over 65,000 bounding boxes. On average, each app contains a maximum of 5.4 screenshots, 305.1 bounding boxes, and 89.2 instructions. The dataset balances fundamental pages (common across apps) and domain-specific pages (unique to each app), ensuring broad coverage of GUI elements and tasks.

To assess meaningful differences between versions, we manually analyzed screenshots to quantify interface changes. Among all screenshots, 74.6\% showed significant differences. We further examined 20 pairs of old and new version screenshots, classifying elements into six categories: layout or icon/text changes, both layout and icon/text changes, additions, deletions, and no change. Results are summarized in the table \ref{tab:diff_2_version}.

\begin{table}[!ht]
\setlength{\belowcaptionskip}{0pt}
    \centering
    \begin{adjustbox}{max width=0.4\textwidth}

   \begin{tabular}{l |c }
    \hline
    \textbf{Element Type} & \textbf{Percentage}\\
    \hline
    Layout Change & 10.0\%\\
    Text/Icon Change & 2.0\%\\
    Layout and Text/Icon Change & 13.6\%\\
    Addition & 30.8\%\\
    Deletion & 14.0\%\\
    No Change & 29.6\%
    \\
    \hline
    \end{tabular}
    \end{adjustbox}

    \caption{Differences between Android old version and Android new version.}
    \label{tab:diff_2_version}
\end{table}


\begin{table*}[htbp]

\setlength{\belowcaptionskip}{0pt}
    \centering
    \scalebox{0.9}{
        \begin{tabular}{l|cc|cc|cc|cc|cc|cc}
            \hline
            \multirow{3}{*}{\textbf{Models}}& 
            \multicolumn{2}{c|}{\multirow{2}{*}{\textbf{General}}}
            & 
            \multicolumn{6}{c|}{\textbf{Android}} &
            \multicolumn{2}{c|}{\multirow{2}{*}{\textbf{iOS}}}& 
            \multicolumn{2}{c}{\multirow{2}{*}{\textbf{Web}}}\\ 
            &\multicolumn{2}{c|}{}&
            \multicolumn{2}{c}{Overall} & 
            \multicolumn{2}{c}{Android Old}&
            \multicolumn{2}{c|}{Android New}&
            \multicolumn{2}{c|}{}
            \\
            \cline{2-13}
            &
            acc$\uparrow$ &dis$\downarrow$&acc$\uparrow$&dis$\downarrow$&acc$\uparrow$&dis$\downarrow$&acc$\uparrow$&dis$\downarrow$&acc$\uparrow$&dis$\downarrow$&acc$\uparrow$&dis$\downarrow$
            \\ \hline
            Cogagent& 72.16 & 14.99 & 75.86 & 14.13 & 76.04 & 13.99 & 75.70 & 14.25 & 68.61 & 16.28 & 66.69 & 15.58\\
            \hline
            SeeClick& 39.90 & 22.72 & 46.63 & 19.27 & 46.86 & 19.07 & 46.42 & 19.45 & 43.57 & 19.71 & 15.37 & 36.96 \\ 
            \hline
            Aria-UI& 77.51 & 9.26 & 81.18 & 8.99 & 80.97 & 9.10 & 81.38 & 8.89 & 77.61 & 9.61 & 66.86 & 9.55 \\ 
            \hline
            OS-Atlas& 81.37 & 8.36 & 84.56 & 8.24 & 84.52 & 8.10 & 84.60 & 8.36 & 79.64 & 8.89 & 74.76 & 7.97 \\ 
            \hline
            UGround& \underline{84.18} & \textbf{7.23} & \underline{87.34} & \textbf{6.89} & \underline{86.94} & \textbf{6.89} & \underline{87.71} & \textbf{6.89} & \underline{82.43} & \textbf{7.42} & \underline{77.62} & \underline{7.94}\\ 
            \hline
            Qwen2.5VL &\textbf{86.43} & \underline{7.72} & \textbf{89.62} & \underline{7.68} & \textbf{88.87} & \underline{7.82} & \textbf{90.29} & \underline{7.55} & \textbf{84.72} & \underline{8.04} & \textbf{79.79} & \textbf{7.35}\\ 
            \hline
        \end{tabular}
    }
    \caption{Accuracy rate (\%) of different LLMs on \texttt{TransBench}.}
    \label{tab:exp_result}
\vspace{-8pt}
\end{table*}


\section{Experiment}
\subsection{Setup}

\paragraph{Models.} We select several top-performing LLMs on ScreenSpot. Specifically, we include Cogagent \cite{hong2024cogagent} (glm-4v \cite{Zeng2024ChatGLMAF}), Seeclick \cite{cheng2024seeclick} (Qwen-VL \cite{Bai2023QwenVLAV}), Aria-UI \cite{yang2024aria} (Aria \cite{Li2024AriaAO}), OS-Atlas \cite{wu2024atlas} (Qwen2-VL \cite{Wang2024Qwen2VLEV}), UGround \cite{gou2024navigating} (Qwen2-VL), and Qwen2.5VL \cite{Qwen2.5-VL}. Among them, UGround and CogAgent have been updated compared to the versions in their paper, and Qwen2.5VL is the latest released model.
\vspace{-3pt}
\paragraph{Implementation Details.} Following \cite{huang2024planning,zhuang2023toolqa, cheng2024seeclick}, we conduct evaluations under two configurations: 1) the Standard Set, where all baselines are tested and evaluated on our complete dataset. Evaluation details can be found in Appendix \ref{appendix2_prompt_details}. 2) the Finetuning Set, which utilizes the rich metadata of our dataset to construct different partitions, allowing models to be fine-tuned on the training sets of each partition and tested on the corresponding test sets. Details of the training set division and finetuning are available in Appendix \ref{appendix2_trainingset_division} and \ref{appendix2_finetuning}.

\vspace{-4pt}
\subsection{Evaluation Metrics}
In line with previous practices, we evaluate \textit{grounding accuracy} on \texttt{TransBench} by determining a prediction to be correct if the predicted location is contained within the ground truth bounding box. Moreover, to compare the precision of click positions at a finer scale, we introduce \textit{an average distance evaluation metric} $D$, which refers to the Euclidean distance between the predicted click position $(x_i, y_i)$ and the center of the Ground Truth bounding box $(\hat{x_i},\hat{y_i})$. To accommodate various interface sizes and ensure readability, we uniformly scale $x, y$ with screen width and height to a range of 0-100 by $x' = \frac{x}{W} \times 100$ and $y' = \frac{y}{H} \times 100$, thereby enabling the comparison of precision across screens with varying width and height.

\begin{equation}
D = \frac{1}{N} \sum_{i=1}^{N} \left\| (\hat{x}'_i, \hat{y}'_i) , (x'_i,y'_i) \right\|
\end{equation}

\subsection{Results}
Table \ref{tab:exp_result} presents the \texttt{TransBench} evaluation results of different LLMs across different platforms and app versions. Several key observations can be made:
\vspace{-3pt}
\paragraph{Overall, Qwen2.5VL achieves the highest accuracy, while UGround minimizes distance.} Qwen2.5VL consistently outperforms all other models in terms of accuracy across all settings, achieving the highest overall accuracy (89.62\%). Meanwhile, UGround exhibits the lowest distance metric in most cases, except on the Web, where Qwen2.5VL attains the smallest distance (7.35). Other models show a notable performance gap compared to these two, with SeeClick and Cogagent demonstrating particularly weak performance, as reflected in both accuracy and distance scores.

\vspace{-3pt}
\paragraph{Models released at different times exhibit varying performance across different versions of Android, with newer models generally performing better on the Android new version.} In detail, older models such as CogAgent (76.04\% old vs 75.70\% new) and SeeClick (46.86\% old vs 46.42\% new) demonstrate better performance on android old version. Conversely, newer models, including Aria-UI, OS-Atlas, UGround, and the top-performing Qwen2.5VL (88.87\% old vs 90.29\% new) achieve higher results on new versions. These results suggest substantial differences between app versions, which potentially come from evolving GUI components and interaction philosophies.

\vspace{-3pt}
\paragraph{GUI agents tend to perform grounding best on Android, followed by iOS, with the worst performance on the web.} As reported in Table \ref{tab:exp_result}, we can find that the performance on Android is always higher than iOS, and the performance of both Android and iOS is substantially better than Web interfaces, no matter which method is chosen. For example, Qwen2.5VL achieves 89.6\%, 84.72\%, and 79.79\% accuracy on Android, iOS, and Web, respectively. The substantially weaker Web results indicate that GUI differences between mobile and web platforms pose a significant challenge, and the observed performance variations between Android and iOS platforms also underscore the inherent heterogeneity within mobile ecosystems. 
\vspace{-3pt}
\paragraph{Our proposed distance serves as a strong complementary evaluation metric to accuracy, offering a finer-grained assessment of precision in grounding tasks.} Unlike accuracy, which simply checks whether a click falls within the bounding box, distance considers the exact position of the click relative to the box's center. It is observed that Cogagent exhibits a large discrepancy between accuracy and distance. Despite only a 7\% accuracy gap compared to Aria-UI, its distance is 62\% higher. Further inspection reveals that Cogagent interprets some tasks as already completed, although we precisely prompt it to perform a click action. Qwen2.5VL achieves the highest accuracy, but it underperforms UGround in terms of the distance metric. With more careful inspection, we speculate that this may be due to its absolute coordinate output (align with screenshot resolution), whereas UGround normalizes coordinates to a 0-1000 scale, making it more robust to varying screen resolutions.

\section{Analysis}

In this section, we fine-tune Aria-UI\footnote{All the code and scripts will be open-sourced. Please see Appendix \ref{appendix:exp} for more details}. We use ARIA-UI as an experimental subject to address three key research questions. \textbf{RQ1:} How does transferability across versions impact the performance of GUI agents, and can fine-tuning on older versions improve adaptability to newer ones? (Sec \ref{sec_1}) \textbf{RQ2:} To what extent can models generalize across platforms (i.e., from Android, iOS, to Web)? (Sec \ref{sec_2}) \textbf{RQ3:} How do models perform when transferring knowledge across applications with varying functionalities, and what are the limitations in cross-application generalization? (Sec \ref{sec_3}) 
\vspace{-2pt}
\subsection{\textit{Cross-Version Transferability} Evaluation}

\begin{table*}[!ht]
\centering
\scalebox{0.85}{
\begin{tabular}{l|cc|cc|cc|cc}
    \hline
    \textbf{Models} & \multicolumn{2}{c|}{\textbf{Android New}} &
    \multicolumn{2}{c|}{\textbf{Android Old}}&
    \multicolumn{2}{c|}{\textbf{iOS}}&
    \multicolumn{2}{c}{\textbf{Web}}\\
    & acc&dis&acc&dis&acc&dis&acc&dis\\
    \hline
    \multicolumn{9}{c}{\cellcolor[HTML]{EFEFEF} \textit{Base Model}} \\ 
    Aria-UI-Base                             &81.38 & 8.89 & 80.97 & 9.10 & 77.61 &9.61 & 66.86 & 9.55\\

    \multicolumn{9}{c}{\cellcolor[HTML]{EFEFEF} \textit{Fine-tuned Model}} \\ 

    Aria-UI-Android-old  & \textbf{88.36} &  \textbf{5.80} &  \textbf{89.37}  & \textbf{5.92}    & \underline{82.57}  & 7.74   & 73.61 & \underline{8.43} \\

    Aria-UI-iOS          & 87.06  & 6.98 & 86.83 & 6.92 &       82.03  & \textbf{7.09}     & \underline{73.66} & 8.74\\


    Aria-UI-General           & \underline{88.15} & \underline{6.17} & \underline{87.20}  & \underline{6.08} &   \textbf{83.15} & \underline{7.21} & \textbf{76.54} & \textbf{7.58}  \\
    \hline
\end{tabular}
}
\caption{Accuracy rate (\%) of Aria-UI after fine-tuning on the Different split of \texttt{TransBench}.}
\label{tab:exp_result1}
\vspace{-3mm}
\end{table*}

    
\label{sec_1}

To investigate the cross-version transferability, we split our dataset into a training set (containing 5,000 samples of low-version Android data) and a test set (composed of high-version Android data, iOS data, and Web data). We then fine-tune the Aria-UI model using the training dataset, resulting in the Aria-UI-Android-old model. We provide the performance of it on the test set in Table \ref{tab:exp_result1}. Training focused on comparing old and new versions demonstrates significant performance improvements across all platforms. Specifically, accuracy on the Android new version increases from 81.38\% to 88.36\%, on iOS from 77.61\% to 82.57\%, and the Web from 66.86\% to 73.61\%. Notably, the performance on Android new and iOS surpasses that of the second-best model, UGround, and approaches the performance of the top-performing Qwen2.5VL. This indicates strong transferability from old to newer versions, suggesting that finetuning on older versions can yield robust performance even after application updates. Furthermore, it highlights the potential of leveraging historical data to improve model adaptability across version changes and even different platforms.

In addition, we evaluated whether fine-tuning data from the old version improves a model’s performance on newly added UI elements in the new version. Using 271 randomly sampled new elements from the updated version, we compared the accuracy of the model before and after fine-tuning on the old Android version.

\begin{table}[h]
\centering
\begin{adjustbox}{max width=0.49\textwidth}
\begin{tabular}{c|c}
\hline
\textbf{Model} & \textbf{New Element Accuracy (\%)} $\uparrow$ \\
\hline
Aria-UI-Base & 80.15 \\
Aria-UI-Android-Low & 87.50 \\
\hline
\end{tabular}
\end{adjustbox}

\caption{Aria-UI Accuracy Change for Newly Added Elements in New Version}

\label{tab:crossversion_error_analysis}
\end{table}

The results in Table \ref{tab:crossversion_error_analysis} indicate that the GUI elements newly added in the new version can benefit from fine-tuning based on the old version.
\vspace{-2pt}
\subsection{\textit{Cross-Platform Transferability} Evaluation}
\label{sec_2}

\begin{table}[t]
\centering
\scalebox{0.6}{
\begin{tabular}{l|cc|cc|cc|cc}
    \hline
    \textbf{Models} & \multicolumn{2}{c|}{\textbf{Android New}} &
    \multicolumn{2}{c|}{\textbf{Android Old}}&
    \multicolumn{2}{c|}{\textbf{iOS}}&
    \multicolumn{2}{c}{\textbf{Web}}\\
    & acc&dis&acc&dis&acc&dis&acc&dis\\
    \hline
    \multicolumn{9}{c}{\cellcolor[HTML]{EFEFEF} \textit{Base Model}} \\ 
    Aria-UI-Base                             &81.38 & 8.89 & 80.97 & 9.10 & 77.61 &9.61 & 66.86 & 9.55\\

    \multicolumn{9}{c}{\cellcolor[HTML]{EFEFEF} \textit{Fine-tuned Model}} \\ 

    Aria-UI-Web                  & 84.87 & 7.70 & 84.08 &  7.70 &   80.62 & 8.39 & 66.49 & 9.60 \\
    \hline
\end{tabular}
}
\caption{Accuracy rate (\%) of Aria-UI after fine-tuning on the Web split of \texttt{TransBench}.}
\label{tab:exp_result2}
 \vspace{-7pt}
\end{table}

To assess cross-platform transferability, we further create two additional training sets: 1) iOS split using 5000 samples of iOS data as the training set (Aria-UI-iOS); and 2) general split, which mixes all available data and randomly select 5000 samples as the training set to maintain consistency (Aria-UI-General). Furthermore, to evaluate transferability from the Web to other platforms, we create a Web split using 4,000 samples of Web data due to the smaller data scale as the training set (Aria-UI-Web). 
\vspace{-3pt}
\paragraph{General, iOS, and Android comparison.} Table \ref{tab:exp_result1} shows the results of Aria-UI-iOS, Aria-UI-Android-old and Aria-UI-General. \textit{It is observed that finetuning on Android data provides the most substantial performance gains across platforms.} For instance, iOS accuracy improves by 4.96\% when fine-tuned on Android data, compared to 4.42\% when fine-tuned on iOS data. This suggests that Android data, being more diverse and representative than iOS, offers better transferability to other platforms, and finetuning on it is highly effective. In addition, \textit{we can found it is not easy to directly transfer from Android or iOS to Web without the web data} since the performance of Aria-UI-iOS and Aria-UI-Android-old is significantly worse than Aria-UI-General.
\vspace{-3pt}
\paragraph{Web results.} Besides that, the result in Table \ref{tab:exp_result2} on the one hand shows that fine-tuning on the Web with 4,000 samples can not significantly enhance performance on the Web test set, but on the other hand, can improve performance on Android and iOS. Combined with our previous findings in Table~\ref{tab:exp_result1}, the Aria-UI-General model, fine-tuned on the general split, achieves the most significant improvement on Web (76.54\%) and iOS (83.15\%). This further confirms that diverse, multi-platform data is crucial for enhancing not only Web performance, but for achieving robust cross-platform transferability.

\subsection{\textit{Cross-Application Transferability} Evaluation}
\label{sec_3}

\begin{figure}[!t]
    \subfigbottomskip=0pt 
    \centering
    \subfigure[Accuracy$\uparrow$]{
    \begin{minipage}[b]{.465\linewidth}
    \includegraphics[width=1.05\linewidth]{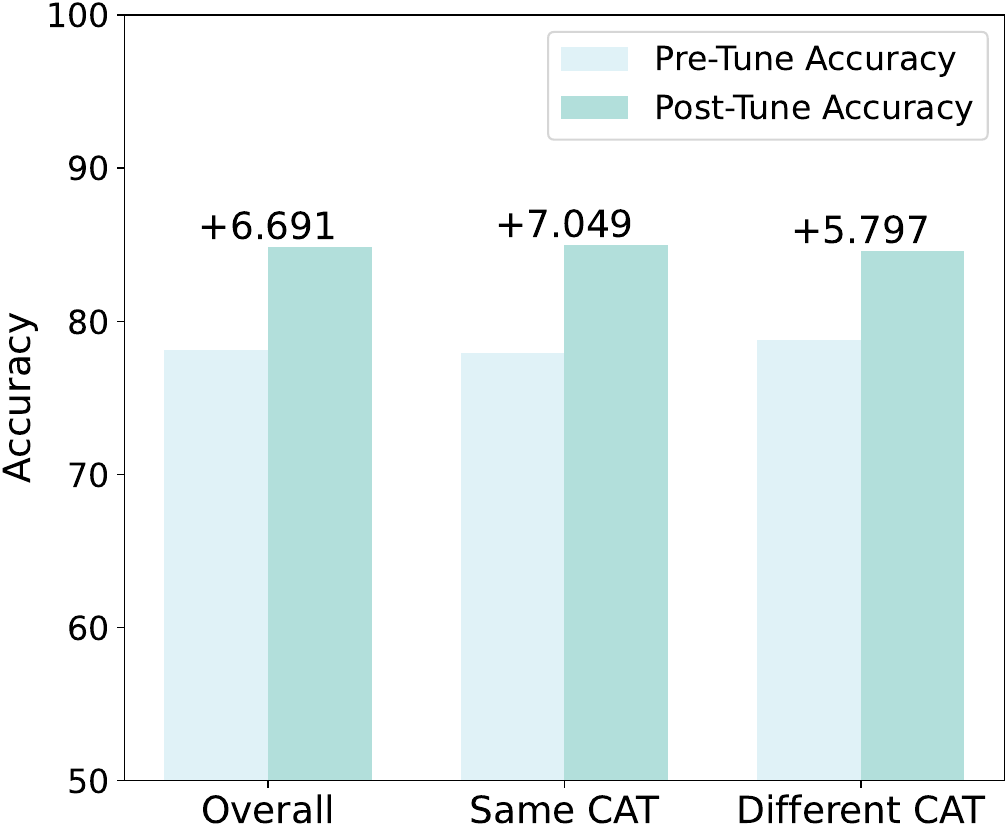}
    \end{minipage}
    }
    \subfigure[Distance$\downarrow$]{
    \begin{minipage}[b]{.465\linewidth}
    \includegraphics[width=1.04\linewidth]{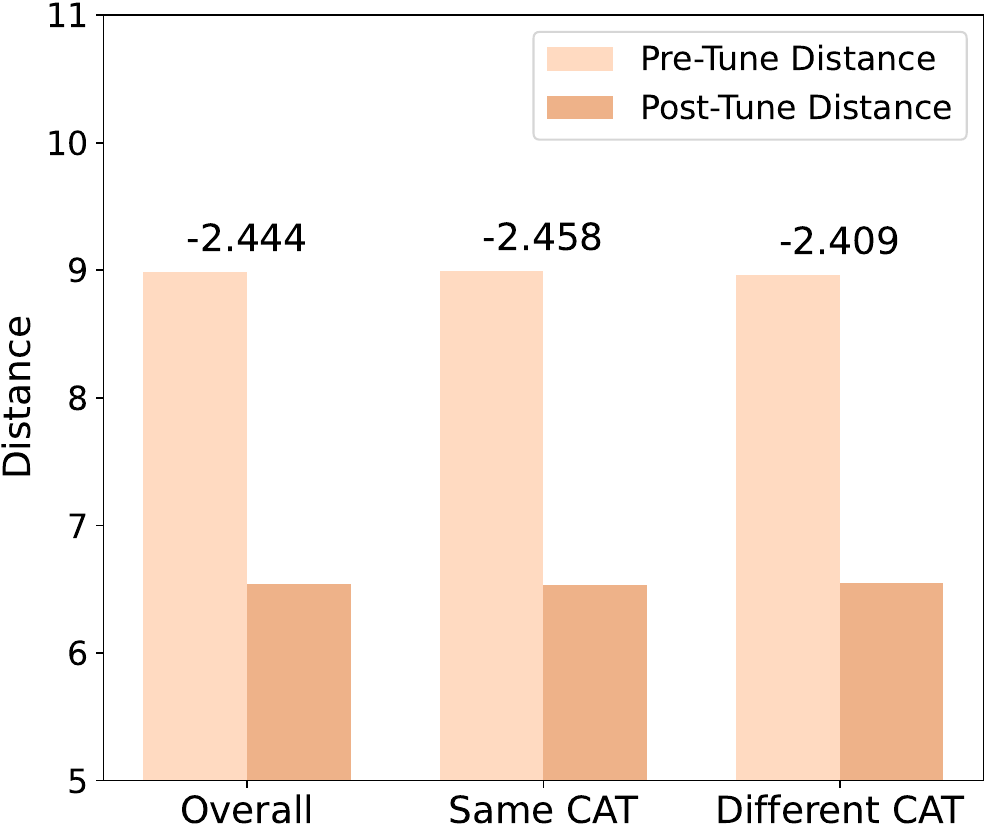}
    \end{minipage}
    } 
    
    \caption{Sub-figure (a), (b) shows the variation of average accuracy and average distance after finetuning Aria-ui on App Split. "CAT" means category. }

    \label{fig:app_finetune}
    \vspace{-9pt}
\end{figure}

To evaluate cross-application transferability, we split app categories into two parts\footnote{The training set has 6247 samples and the test set has 15991 samples. We randomly selected 5000 samples from the training set for finetuning.}: 1) the training set contains 18 Apps from the first seven categories; and 2) the test set, which is composed of 35 Apps from the same seven categories (i.e., Same CAT), and 28 Apps from other categories (i.e, Different CAT). Then we fine-tune the Aria-UI model on the training set and inferences at these two different test sets. Figure \ref{fig:app_finetune} shows the significant improvements in accuracy and reductions in the distance across both apps in \textit{same category} and \textit{different category}. However, the performance gains are slightly more pronounced for apps in same category, benefiting from the task similarity. Despite this, the difference between the two category types is not substantial, indicating that app similarity has less impact on transferability compared to platform and version differences. This suggests that while app-specific finetuning can yield improvements, the overall transferability of GUI agents is more influenced by platform and version adaptability.

\section{Conclusion}

In this paper, we explore three major fine-grained aspects of transferability (i.e., \textit{cross-version, cross-platform, and cross-application}) of grounding capabilities for GUI agents to better accommodate diverse user instructions and complex real-world scenarios. To this end, we build the first comprehensive benchmark -- \texttt{TransBench}, covering various applications spanning different versions and platforms. Our experimental results on a more fine-grained evaluation showcase that there is still a big gap between different levels of transferability, and we hope our benchmarks and new metrics can pave the way for more effective and adaptable GUI agents in practical applications.


\section*{Limitation}
One notable limitation of our approach is the high computational requirements for training and fine-tuning the models. The extensive dataset, combined with the need for multi-dimensional partitioning and rigorous evaluation of transferability across versions, platforms, and applications, demands significant computational resources. This can pose challenges for researchers or organizations with limited access to high-performance computing infrastructure, potentially restricting the reproducibility and scalability of our methods. 

\section*{Ethical Statement}
This study adheres to ethical guidelines by ensuring the security and privacy of all data used during research. Screenshots collected from real applications are anonymized, and no sensitive user information is included. Additionally, manual annotations are performed under strict supervision to avoid human bias in data labeling. AI assistants, including DeepSeek and Qwen, are used to assist in understanding code and translating. Open-source components are used responsibly, with proper attribution given to their developers. The benchmark dataset will be shared following data-sharing agreements, promoting transparency while safeguarding any proprietary information, and ensuring responsible dissemination and compliance with ethical standards in academic research.

\section*{Acknowledgments}
Thanks for the insightful comments and feedback from the reviewers. This work was supported by the National Natural Science Foundation of China (No. 62406015), and CCF-Baidu Open Fund (No. CCF-BAIDU202411).

\bibliography{guiagent}

\begin{thebibliography}{43}
\providecommand{\natexlab}[1]{#1}

\bibitem[{Baechler et~al.(2024)Baechler, Sunkara, Wang, Zubach, Mansoor, Etter, Carbune, Lin, Chen, and Sharma}]{ijcai2024p339}
Gilles Baechler, Srinivas Sunkara, Maria Wang, Fedir Zubach, Hassan Mansoor, Vincent Etter, Victor Carbune, Jason Lin, Jindong Chen, and Abhanshu Sharma. 2024.
\newblock \href {https://doi.org/10.24963/ijcai.2024/339} {Screenai: A vision-language model for ui and infographics understanding}.
\newblock In \emph{Proceedings of the Thirty-Third International Joint Conference on Artificial Intelligence, {IJCAI-24}}, pages 3058--3068. International Joint Conferences on Artificial Intelligence Organization.
\newblock Main Track.

\bibitem[{Bai et~al.(2021)Bai, Zang, Xu, Sunkara, Rastogi, Chen, and y~Arcas}]{Bai2021UIBertLG}
Chongyang Bai, Xiaoxue Zang, Ying Xu, Srinivas Sunkara, Abhinav Rastogi, Jindong Chen, and Blaise~Ag{\"u}era y~Arcas. 2021.
\newblock \href {https://api.semanticscholar.org/CorpusID:236493482} {Uibert: Learning generic multimodal representations for ui understanding}.
\newblock In \emph{International Joint Conference on Artificial Intelligence}.

\bibitem[{Bai et~al.(2023)Bai, Bai, Yang, Wang, Tan, Wang, Lin, Zhou, and Zhou}]{Bai2023QwenVLAV}
Jinze Bai, Shuai Bai, Shusheng Yang, Shijie Wang, Sinan Tan, Peng Wang, Junyang Lin, Chang Zhou, and Jingren Zhou. 2023.
\newblock \href {https://api.semanticscholar.org/CorpusID:261101015} {Qwen-vl: A versatile vision-language model for understanding, localization, text reading, and beyond}.

\bibitem[{Burns et~al.(2022)Burns, Arsan, Agrawal, Kumar, Saenko, and Plummer}]{hwang2025motif}
Andrea Burns, Deniz Arsan, Sanjna Agrawal, Ranjitha Kumar, Kate Saenko, and Bryan~A. Plummer. 2022.
\newblock A dataset for interactive vision language navigation with unknown command feasibility.
\newblock In \emph{European Conference on Computer Vision (ECCV)}.

\bibitem[{Chen et~al.(2024{\natexlab{a}})Chen, Huang, Wu, Tang, Chen, Bai, He, Wang, Zhou, Li, Zhou, Yu, Gao, Zhang, Gui, Li, Wan, Zhou, Gao, and Sun}]{Chen2024GUIWORLDAD}
Dongping Chen, Yue Huang, Siyuan Wu, Jingyu Tang, Liuyi Chen, Yilin Bai, Zhigang He, Chenlong Wang, Huichi Zhou, Yiqiang Li, Tianshuo Zhou, Yue Yu, Chujie Gao, Qihui Zhang, Yi~Gui, Zhen Li, Yao Wan, Pan Zhou, Jianfeng Gao, and Lichao Sun. 2024{\natexlab{a}}.
\newblock \href {https://api.semanticscholar.org/CorpusID:270560712} {Gui-world: A dataset for gui-oriented multimodal llm-based agents}.
\newblock \emph{ArXiv}, abs/2406.10819.

\bibitem[{Chen et~al.(2024{\natexlab{b}})Chen, Cui, Hu, Qin, Fang, Zhao, Wang, Liu, Chen, Huo et~al.}]{chen2024guicourse}
Wentong Chen, Junbo Cui, Jinyi Hu, Yujia Qin, Junjie Fang, Yue Zhao, Chongyi Wang, Jun Liu, Guirong Chen, Yupeng Huo, et~al. 2024{\natexlab{b}}.
\newblock Guicourse: From general vision language models to versatile gui agents.
\newblock \emph{arXiv preprint arXiv:2406.11317}.

\bibitem[{Cheng et~al.(2024)Cheng, Sun, Chu, Xu, Li, Zhang, and Wu}]{cheng2024seeclick}
Kanzhi Cheng, Qiushi Sun, Yougang Chu, Fangzhi Xu, Yantao Li, Jianbing Zhang, and Zhiyong Wu. 2024.
\newblock Seeclick: Harnessing gui grounding for advanced visual gui agents.
\newblock \emph{arXiv preprint arXiv:2401.10935}.

\bibitem[{Chu et~al.(2023)Chu, Qiao, Lin, Xu, Yang, Hu, Wei, Zhang, Zhang, Wei et~al.}]{chu2023mobilevlm}
Xiangxiang Chu, Limeng Qiao, Xinyang Lin, Shuang Xu, Yang Yang, Yiming Hu, Fei Wei, Xinyu Zhang, Bo~Zhang, Xiaolin Wei, et~al. 2023.
\newblock Mobilevlm: A fast, strong and open vision language assistant for mobile devices.
\newblock \emph{arXiv preprint arXiv:2312.16886}.

\bibitem[{Deng et~al.(2024)Deng, Gu, Zheng, Chen, Stevens, Wang, Sun, and Su}]{deng2024mind2web}
Xiang Deng, Yu~Gu, Boyuan Zheng, Shijie Chen, Sam Stevens, Boshi Wang, Huan Sun, and Yu~Su. 2024.
\newblock Mind2web: Towards a generalist agent for the web.
\newblock \emph{Advances in Neural Information Processing Systems}, 36.

\bibitem[{Gao et~al.(2024{\natexlab{a}})Gao, Ji, Bai, Ouyang, Li, Mao, Wu, Zhang, Wang, Guo et~al.}]{gao2024assistgui}
Difei Gao, Lei Ji, Zechen Bai, Mingyu Ouyang, Peiran Li, Dongxing Mao, Qinchen Wu, Weichen Zhang, Peiyi Wang, Xiangwu Guo, et~al. 2024{\natexlab{a}}.
\newblock Assistgui: Task-oriented pc graphical user interface automation.
\newblock In \emph{Proceedings of the IEEE/CVF Conference on Computer Vision and Pattern Recognition}, pages 13289--13298.

\bibitem[{Gao et~al.(2024{\natexlab{b}})Gao, Zhang, Wang, Wang, Li, and Xu}]{gao2024mobileviews}
Longxi Gao, Li~Zhang, Shihe Wang, Shangguang Wang, Yuanchun Li, and Mengwei Xu. 2024{\natexlab{b}}.
\newblock Mobileviews: A large-scale mobile gui dataset.
\newblock \emph{arXiv preprint arXiv:2409.14337}.

\bibitem[{Gou et~al.(2024)Gou, Wang, Zheng, Xie, Chang, Shu, Sun, and Su}]{gou2024navigating}
Boyu Gou, Ruohan Wang, Boyuan Zheng, Yanan Xie, Cheng Chang, Yiheng Shu, Huan Sun, and Yu~Su. 2024.
\newblock Navigating the digital world as humans do: Universal visual grounding for gui agents.
\newblock \emph{arXiv preprint arXiv:2410.05243}.

\bibitem[{Hong et~al.(2024)Hong, Wang, Lv, Xu, Yu, Ji, Wang, Wang, Dong, Ding et~al.}]{hong2024cogagent}
Wenyi Hong, Weihan Wang, Qingsong Lv, Jiazheng Xu, Wenmeng Yu, Junhui Ji, Yan Wang, Zihan Wang, Yuxiao Dong, Ming Ding, et~al. 2024.
\newblock Cogagent: A visual language model for gui agents.
\newblock In \emph{Proceedings of the IEEE/CVF Conference on Computer Vision and Pattern Recognition}, pages 14281--14290.

\bibitem[{Hu et~al.(2024)Hu, Wang, Wang, Zhang, Guo, Chen, Wang, and Zhou}]{Hu2024AUITestAgentAR}
Yongxiang Hu, Xuan Wang, Yingchuan Wang, Yu~Zhang, Shiyu Guo, Chaoyi Chen, Xin Wang, and Yangfan Zhou. 2024.
\newblock \href {https://api.semanticscholar.org/CorpusID:271162246} {Auitestagent: Automatic requirements oriented gui function testing}.
\newblock \emph{ArXiv}, abs/2407.09018.

\bibitem[{Huang et~al.(2024)Huang, Zhong, Lu, Zhu, Gao, Liu, Hou, Zeng, Wang, Shang et~al.}]{huang2024planning}
Shijue Huang, Wanjun Zhong, Jianqiao Lu, Qi~Zhu, Jiahui Gao, Weiwen Liu, Yutai Hou, Xingshan Zeng, Yasheng Wang, Lifeng Shang, et~al. 2024.
\newblock Planning, creation, usage: Benchmarking llms for comprehensive tool utilization in real-world complex scenarios.
\newblock \emph{arXiv preprint arXiv:2401.17167}.

\bibitem[{Kapoor et~al.(2025)Kapoor, Butala, Russak, Koh, Kamble, AlShikh, and Salakhutdinov}]{kapoor2025omniact}
Raghav Kapoor, Yash~Parag Butala, Melisa Russak, Jing~Yu Koh, Kiran Kamble, Waseem AlShikh, and Ruslan Salakhutdinov. 2025.
\newblock Omniact: A dataset and benchmark for enabling multimodal generalist autonomous agents for desktop and web.
\newblock In \emph{European Conference on Computer Vision}, pages 161--178. Springer.

\bibitem[{Li et~al.(2024)Li, Liu, Wu, Wang, Shen, Qu, Niu, Wang, Chen, and Li}]{Li2024AriaAO}
Dongxu Li, Yudong Liu, Haoning Wu, Yue Wang, Zhiqi Shen, Bowen Qu, Xinyao Niu, Guoyin Wang, Bei Chen, and Junnan Li. 2024.
\newblock \href {https://api.semanticscholar.org/CorpusID:273229053} {Aria: An open multimodal native mixture-of-experts model}.
\newblock \emph{ArXiv}, abs/2410.05993.

\bibitem[{Li et~al.(2020)Li, He, Zhou, Zhang, and Baldridge}]{li2020mapping}
Yang Li, Jiacong He, Xin Zhou, Yuan Zhang, and Jason Baldridge. 2020.
\newblock Mapping natural language instructions to mobile ui action sequences.
\newblock \emph{arXiv preprint arXiv:2005.03776}.

\bibitem[{Liu et~al.(2024{\natexlab{a}})Liu, Qin, Liang, Dong, Lai, Zhang, Zhao, Iong, Sun, Wang et~al.}]{liu2024autoglm}
Xiao Liu, Bo~Qin, Dongzhu Liang, Guang Dong, Hanyu Lai, Hanchen Zhang, Hanlin Zhao, Iat~Long Iong, Jiadai Sun, Jiaqi Wang, et~al. 2024{\natexlab{a}}.
\newblock Autoglm: Autonomous foundation agents for guis.
\newblock \emph{arXiv preprint arXiv:2411.00820}.

\bibitem[{Liu et~al.(2023)Liu, Yu, Zhang, Xu, Lei, Lai, Gu, Ding, Men, Yang et~al.}]{liu2023agentbench}
Xiao Liu, Hao Yu, Hanchen Zhang, Yifan Xu, Xuanyu Lei, Hanyu Lai, Yu~Gu, Hangliang Ding, Kaiwen Men, Kejuan Yang, et~al. 2023.
\newblock Agentbench: Evaluating llms as agents.
\newblock \emph{arXiv preprint arXiv:2308.03688}.

\bibitem[{Liu et~al.(2024{\natexlab{b}})Liu, Zhang, Gu, Iong, Xu, Song, Zhang, Lai, Liu, Zhao, Sun, Yang, Yang, Qi, Yao, Sun, Cheng, Zheng, Yu, Zhang, Hong, Ding, Pan, Gu, Zeng, Du, Song, Su, Dong, and Tang}]{Liu2024VisualAgentBenchTL}
Xiao Liu, Tianjie Zhang, Yu~Gu, Iat~Long Iong, Yifan Xu, Xixuan Song, Shudan Zhang, Hanyu Lai, Xinyi Liu, Hanlin Zhao, Jiadai Sun, Xinyue Yang, Yu~Yang, Zehan Qi, Shuntian Yao, Xueqiao Sun, Siyi Cheng, Qi~Zheng, Hao Yu, Hanchen Zhang, Wenyi Hong, Ming Ding, Lihang Pan, Xiaotao Gu, Aohan Zeng, Zhengxiao Du, Chan~Hee Song, Yu~Su, Yuxiao Dong, and Jie Tang. 2024{\natexlab{b}}.
\newblock \href {https://api.semanticscholar.org/CorpusID:271854812} {Visualagentbench: Towards large multimodal models as visual foundation agents}.
\newblock \emph{ArXiv}, abs/2408.06327.

\bibitem[{Liu et~al.(2024{\natexlab{c}})Liu, Zhang, Gu, Iong, Xu, Song, Zhang, Lai, Liu, Zhao et~al.}]{liu2024visualagentbench}
Xiao Liu, Tianjie Zhang, Yu~Gu, Iat~Long Iong, Yifan Xu, Xixuan Song, Shudan Zhang, Hanyu Lai, Xinyi Liu, Hanlin Zhao, et~al. 2024{\natexlab{c}}.
\newblock Visualagentbench: Towards large multimodal models as visual foundation agents.
\newblock \emph{arXiv preprint arXiv:2408.06327}.

\bibitem[{Liu et~al.(2020)Liu, Wang, Niu, Wu, Che, and Liu}]{liu-etal-2020-towards-conversational}
Zeming Liu, Haifeng Wang, Zheng-Yu Niu, Hua Wu, Wanxiang Che, and Ting Liu. 2020.
\newblock \href {https://doi.org/10.18653/v1/2020.acl-main.98} {Towards conversational recommendation over multi-type dialogs}.
\newblock In \emph{Proceedings of the 58th Annual Meeting of the Association for Computational Linguistics}, pages 1036--1049, Online. Association for Computational Linguistics.

\bibitem[{Lu et~al.(2024{\natexlab{a}})Lu, Shao, Liu, Meng, Li, Chen, Huang, Zhang, Qiao, and Luo}]{lu2024gui}
Quanfeng Lu, Wenqi Shao, Zitao Liu, Fanqing Meng, Boxuan Li, Botong Chen, Siyuan Huang, Kaipeng Zhang, Yu~Qiao, and Ping Luo. 2024{\natexlab{a}}.
\newblock Gui odyssey: A comprehensive dataset for cross-app gui navigation on mobile devices.
\newblock \emph{arXiv preprint arXiv:2406.08451}.

\bibitem[{Lu et~al.(2024{\natexlab{b}})Lu, Yang, Shen, and Awadallah}]{Lu2024OmniParserFP}
Yadong Lu, Jianwei Yang, Yelong Shen, and Ahmed Awadallah. 2024{\natexlab{b}}.
\newblock \href {https://api.semanticscholar.org/CorpusID:271601072} {Omniparser for pure vision based gui agent}.
\newblock \emph{ArXiv}, abs/2408.00203.

\bibitem[{Ma et~al.(2024)Ma, Zhang, and Zhao}]{ma-etal-2024-coco}
Xinbei Ma, Zhuosheng Zhang, and Hai Zhao. 2024.
\newblock \href {https://doi.org/10.18653/v1/2024.findings-acl.539} {{C}o{C}o-agent: A comprehensive cognitive {MLLM} agent for smartphone {GUI} automation}.
\newblock In \emph{Findings of the Association for Computational Linguistics: ACL 2024}, pages 9097--9110, Bangkok, Thailand. Association for Computational Linguistics.

\bibitem[{Mukhtar(2025)}]{mukhtar2025artificial}
Hamid Mukhtar. 2025.
\newblock Artificial intelligence techniques for human-machine interaction.
\newblock In \emph{Artificial Intelligence and Multimodal Signal Processing in Human-Machine Interaction}, pages 19--42. Elsevier.

\bibitem[{Shi et~al.(2023)Shi, Liu, Wang, Leng, Xue, Zhang, and Zhang}]{shi-etal-2023-midmed}
Xiaoming Shi, Zeming Liu, Chuan Wang, Haitao Leng, Kui Xue, Xiaofan Zhang, and Shaoting Zhang. 2023.
\newblock \href {https://doi.org/10.18653/v1/2023.acl-long.453} {{M}id{M}ed: Towards mixed-type dialogues for medical consultation}.
\newblock In \emph{Proceedings of the 61st Annual Meeting of the Association for Computational Linguistics (Volume 1: Long Papers)}, pages 8145--8157, Toronto, Canada. Association for Computational Linguistics.

\bibitem[{Team(2025)}]{Qwen2.5-VL}
Qwen Team. 2025.
\newblock \href {https://qwenlm.github.io/blog/qwen2.5-vl/} {Qwen2.5-vl}.

\bibitem[{Wang et~al.(2024{\natexlab{a}})Wang, Qin, Lin, Pan, and Wong}]{tool_tut}
Hongru Wang, Yujia Qin, Yankai Lin, Jeff~Z. Pan, and Kam-Fai Wong. 2024{\natexlab{a}}.
\newblock \href {https://doi.org/10.1145/3626772.3661381} {Empowering large language models: Tool learning for real-world interaction}.
\newblock In \emph{Proceedings of the 47th International ACM SIGIR Conference on Research and Development in Information Retrieval}, SIGIR '24, page 2983–2986, New York, NY, USA. Association for Computing Machinery.

\bibitem[{Wang et~al.(2024{\natexlab{b}})Wang, Xu, Jia, Zhang, Yan, Shen, Zhang, Huang, and Sang}]{wang2024mobile}
Junyang Wang, Haiyang Xu, Haitao Jia, Xi~Zhang, Ming Yan, Weizhou Shen, Ji~Zhang, Fei Huang, and Jitao Sang. 2024{\natexlab{b}}.
\newblock Mobile-agent-v2: Mobile device operation assistant with effective navigation via multi-agent collaboration.
\newblock \emph{arXiv preprint arXiv:2406.01014}.

\bibitem[{Wang et~al.(2024{\natexlab{c}})Wang, Xia, Gu, Zhao, Shen, Meng, Wang, and Xu}]{wang2024ant}
Ke~Wang, Tianyu Xia, Zhangxuan Gu, Yi~Zhao, Shuheng Shen, Changhua Meng, Weiqiang Wang, and Ke~Xu. 2024{\natexlab{c}}.
\newblock E-ant: A large-scale dataset for efficient automatic gui navigation.
\newblock \emph{arXiv preprint arXiv:2406.14250}.

\bibitem[{Wang et~al.(2024{\natexlab{d}})Wang, Zhu, Ren, Liu, Li, Zhang, Zhang, Wu, Zhan, Liu, and Wang}]{wang_survey_2024}
Ke~Wang, Jiahui Zhu, Minjie Ren, Zeming Liu, Shiwei Li, Zongye Zhang, Chenkai Zhang, Xiaoyu Wu, Qiqi Zhan, Qingjie Liu, and Yunhong Wang. 2024{\natexlab{d}}.
\newblock \href {http://arxiv.org/abs/2410.12896} {A survey on data synthesis and augmentation for large language models}.
\newblock \emph{ArXiv}, abs/2410.12896.

\bibitem[{Wang et~al.(2024{\natexlab{e}})Wang, Bai, Tan, Wang, Fan, Bai, Chen, Liu, Wang, Ge, Fan, Dang, Du, Ren, Men, Liu, Zhou, Zhou, and Lin}]{Wang2024Qwen2VLEV}
Peng Wang, Shuai Bai, Sinan Tan, Shijie Wang, Zhihao Fan, Jinze Bai, Ke-Yang Chen, Xuejing Liu, Jialin Wang, Wenbin Ge, Yang Fan, Kai Dang, Mengfei Du, Xuancheng Ren, Rui Men, Dayiheng Liu, Chang Zhou, Jingren Zhou, and Junyang Lin. 2024{\natexlab{e}}.
\newblock \href {https://api.semanticscholar.org/CorpusID:272704132} {Qwen2-vl: Enhancing vision-language model's perception of the world at any resolution}.
\newblock \emph{ArXiv}, abs/2409.12191.

\bibitem[{Wang et~al.(2024{\natexlab{f}})Wang, Liu, Chen, Gan, Zeng, Yu, Hao, Shao, Wang, and Tang}]{wang2024gui}
Shuai Wang, Weiwen Liu, Jingxuan Chen, Weinan Gan, Xingshan Zeng, Shuai Yu, Xinlong Hao, Kun Shao, Yasheng Wang, and Ruiming Tang. 2024{\natexlab{f}}.
\newblock Gui agents with foundation models: A comprehensive survey.
\newblock \emph{arXiv preprint arXiv:2411.04890}.

\bibitem[{Wu et~al.(2024{\natexlab{a}})Wu, Xu, Liu, Tan, Liu, Li, Luan, Wang, and Shang}]{wu2024mobilevlm}
Qinzhuo Wu, Weikai Xu, Wei Liu, Tao Tan, Jianfeng Liu, Ang Li, Jian Luan, Bin Wang, and Shuo Shang. 2024{\natexlab{a}}.
\newblock Mobilevlm: A vision-language model for better intra-and inter-ui understanding.
\newblock \emph{arXiv preprint arXiv:2409.14818}.

\bibitem[{Wu et~al.(2024{\natexlab{b}})Wu, Xu, Liu, Tan, Liujianfeng, Li, Luan, Wang, and Shang}]{wu-etal-2024-mobilevlm}
Qinzhuo Wu, Weikai Xu, Wei Liu, Tao Tan, Liujian Liujianfeng, Ang Li, Jian Luan, Bin Wang, and Shuo Shang. 2024{\natexlab{b}}.
\newblock \href {https://doi.org/10.18653/v1/2024.findings-emnlp.599} {{M}obile{VLM}: A vision-language model for better intra- and inter-{UI} understanding}.
\newblock In \emph{Findings of the Association for Computational Linguistics: EMNLP 2024}, pages 10231--10251, Miami, Florida, USA. Association for Computational Linguistics.

\bibitem[{Wu et~al.(2024{\natexlab{c}})Wu, Wu, Xu, Wang, Sun, Jia, Cheng, Ding, Chen, Liang et~al.}]{wu2024atlas}
Zhiyong Wu, Zhenyu Wu, Fangzhi Xu, Yian Wang, Qiushi Sun, Chengyou Jia, Kanzhi Cheng, Zichen Ding, Liheng Chen, Paul~Pu Liang, et~al. 2024{\natexlab{c}}.
\newblock Os-atlas: A foundation action model for generalist gui agents.
\newblock \emph{arXiv preprint arXiv:2410.23218}.

\bibitem[{Yang et~al.(2023)Yang, Yue, and He}]{yang2023auto}
Hui Yang, Sifu Yue, and Yunzhong He. 2023.
\newblock Auto-gpt for online decision making: Benchmarks and additional opinions.
\newblock \emph{arXiv preprint arXiv:2306.02224}.

\bibitem[{Yang et~al.(2024)Yang, Wang, Li, Luo, Chen, Huang, and Li}]{yang2024aria}
Yuhao Yang, Yue Wang, Dongxu Li, Ziyang Luo, Bei Chen, Chao Huang, and Junnan Li. 2024.
\newblock Aria-ui: Visual grounding for gui instructions.
\newblock \emph{arXiv preprint arXiv:2412.16256}.

\bibitem[{Zeng et~al.(2024)Zeng, Xu, Wang, Zhang, Yin, Rojas, Feng, Zhao, Lai, Yu, Wang, Sun, Zhang, Cheng, Gui, Tang, Zhang, Li, Zhao, Wu, Zhong, yue Liu, Huang, Zhang, Zheng, Lu, Duan, Zhang, Cao, Yang, Tam, Zhao, Liu, Xia, Zhang, Gu, Lv, Liu, Liu, Yang, Song, Zhang, An, Xu, Niu, Yang, Li, Bai, Dong, Qi, Wang, Yang, Du, Hou, and Wang}]{Zeng2024ChatGLMAF}
Team Glm~Aohan Zeng, Bin Xu, Bowen Wang, Chenhui Zhang, Da~Yin, Diego Rojas, Guanyu Feng, Hanlin Zhao, Hanyu Lai, Hao Yu, Hongning Wang, Jiadai Sun, Jiajie Zhang, Jiale Cheng, Jiayi Gui, Jie Tang, Jing Zhang, Juanzi Li, Lei Zhao, Lindong Wu, Lucen Zhong, Ming yue Liu, Minlie Huang, Peng Zhang, Qinkai Zheng, Rui Lu, Shuaiqi Duan, Shudan Zhang, Shulin Cao, Shuxun Yang, Weng~Lam Tam, Wenyi Zhao, Xiao Liu, Xiaoyu Xia, Xiaohan Zhang, Xiaotao Gu, Xin Lv, Xinghan Liu, Xinyi Liu, Xinyue Yang, Xixuan Song, Xunkai Zhang, Yi~An, Yifan Xu, Yilin Niu, Yuantao Yang, Yueyan Li, Yushi Bai, Yuxiao Dong, Zehan Qi, Zhaoyu Wang, Zhenyi Yang, Zhengxiao Du, Zhen-Ping Hou, and Zihan Wang. 2024.
\newblock \href {https://api.semanticscholar.org/CorpusID:270562306} {Chatglm: A family of large language models from glm-130b to glm-4 all tools}.
\newblock \emph{ArXiv}, abs/2406.12793.

\bibitem[{Zhang et~al.(2024)Zhang, He, Qian, Li, Li, Qin, Kang, Ma, Lin, Rajmohan, Zhang, and Zhang}]{Zhang2024LargeLM}
Chaoyun Zhang, Shilin He, Jiaxu Qian, Bowen Li, Liqun Li, Si~Qin, Yu~Kang, Ming-Jie Ma, Qingwei Lin, S.~Rajmohan, Dongmei Zhang, and Qi~Zhang. 2024.
\newblock \href {https://api.semanticscholar.org/CorpusID:274306375} {Large language model-brained gui agents: A survey}.
\newblock \emph{ArXiv}, abs/2411.18279.

\bibitem[{Zhuang et~al.(2023)Zhuang, Yu, Wang, Sun, and Zhang}]{zhuang2023toolqa}
Yuchen Zhuang, Yue Yu, Kuan Wang, Haotian Sun, and Chao Zhang. 2023.
\newblock Toolqa: A dataset for llm question answering with external tools.
\newblock \emph{Advances in Neural Information Processing Systems}, 36:50117--50143.

\end{thebibliography}

\clearpage
\appendix

\section{Data Collection}
\label{appendix:data_collection}

\subsection{Prompt Details}
\label{appendix1_prompt_details}

The prompts we used are shown in Tabel \ref{appendix:data_collection_1}, \ref{appendix:data_collection_2} and \ref{appendix:data_collection_3}. 

\vspace{-7pt}

\begin{table}[!ht]
\small
    \centering
    \colorbox{blue!8}{
    \begin{tabular}{@{}p{7.2cm}}
    You are a powerful GUI recognizer, your mission is to accurately recognize GUI element on screenshot and output information.\\
\\
You have two input images, first image is a crop of your target GUI element, second is a target GUI element zoomed-in view of full screenshot.\\
\\
Output three types of information, including: Visual information such as "The vertical three-dot button, with text ’more‘ at the bottom", Positional information such as "Next to the entry "28 YEARS LATER - Official Trailer", and the Functional information such as "Access more options for the video entry". The output should be in chinese and included in json dict: \{"\textcolor{myred}{Visual}":"visual information", "\textcolor{myred}{Positional}": "positional information", "\textcolor{myred}{Functional}": "functional information"\}.\\\end{tabular}
    }
    \caption{The prompt used to generate the bounding box attributes, including visual, positional, and functional information.}
    \label{appendix:data_collection_1}
\end{table}

\vspace{-17pt}
\begin{table}[h]
\small
    \centering
    \colorbox{blue!8}{
        \begin{tabular}{@{}p{7.2cm}}
            
            You are a GUI annotator. The input screenshot is the \textcolor{myred}{\itshape \{page\_title\} \upshape} page of the \textcolor{myred}{\itshape \{app\_name\} \upshape} application. Please fully describe the page in Chinese.\end{tabular}
    }
    \caption{The prompts to generate the screen summaries using information of app names with page titles.}
    \label{appendix:data_collection_2}
\end{table}

\vspace{-17pt}
\begin{table}[!h]
\small
    \centering
    \colorbox{blue!8}{
    \begin{tabular}{@{}p{7.2cm}}
    Your mission is to generate instructions that correspond to potential interactions when user want to act with the specified GUI element on current screen, such as instruction "watch video about happy dog." corresponding to a video about a happy dog in video list on the screen. Remember that your output should like a normal user instruction.\\\\Your output based on three part of input information, first is a screen summary, second is the description of target GUI element, third is the current app name and page name. Input information may conflict or including some errors, neglect these conflicts or errors, ensure that the instructions you generate correspond uniquely to the GUI on the screen. \\\\Input1: Screen summary is \{ "screen\_summary": "\textcolor{myred}{\itshape\{screen\_summary\}\upshape}"\}.\\
Input2: The target GUI element is \textcolor{myred}{\itshape\{caption\_data\} \upshape}.\\
    Input3: The current app is \textcolor{myred}{\itshape\{app\_name\}\upshape} app, current page is \textcolor{myred}{\itshape\{page\_name\}\upshape} page.\\\\
    The output should be in Chinese and included in json dict: \{"Instruction": Instructions that the user would say.\}\end{tabular}
    }
    
    \caption{The prompts to generate the grounding instructions using information of screen summary, description of target GUI element, and app names with page titles. Caption data is a json dict like: \{"Visual": "xxx", "Positional": "xxx", "Functional": "xxx"\}}
    \label{appendix:data_collection_3}
\end{table}

\subsection{ Example Description}
\label{appendix:example}

\begin{figure}[!h]
    \centering
    \includegraphics[width=.99\linewidth]{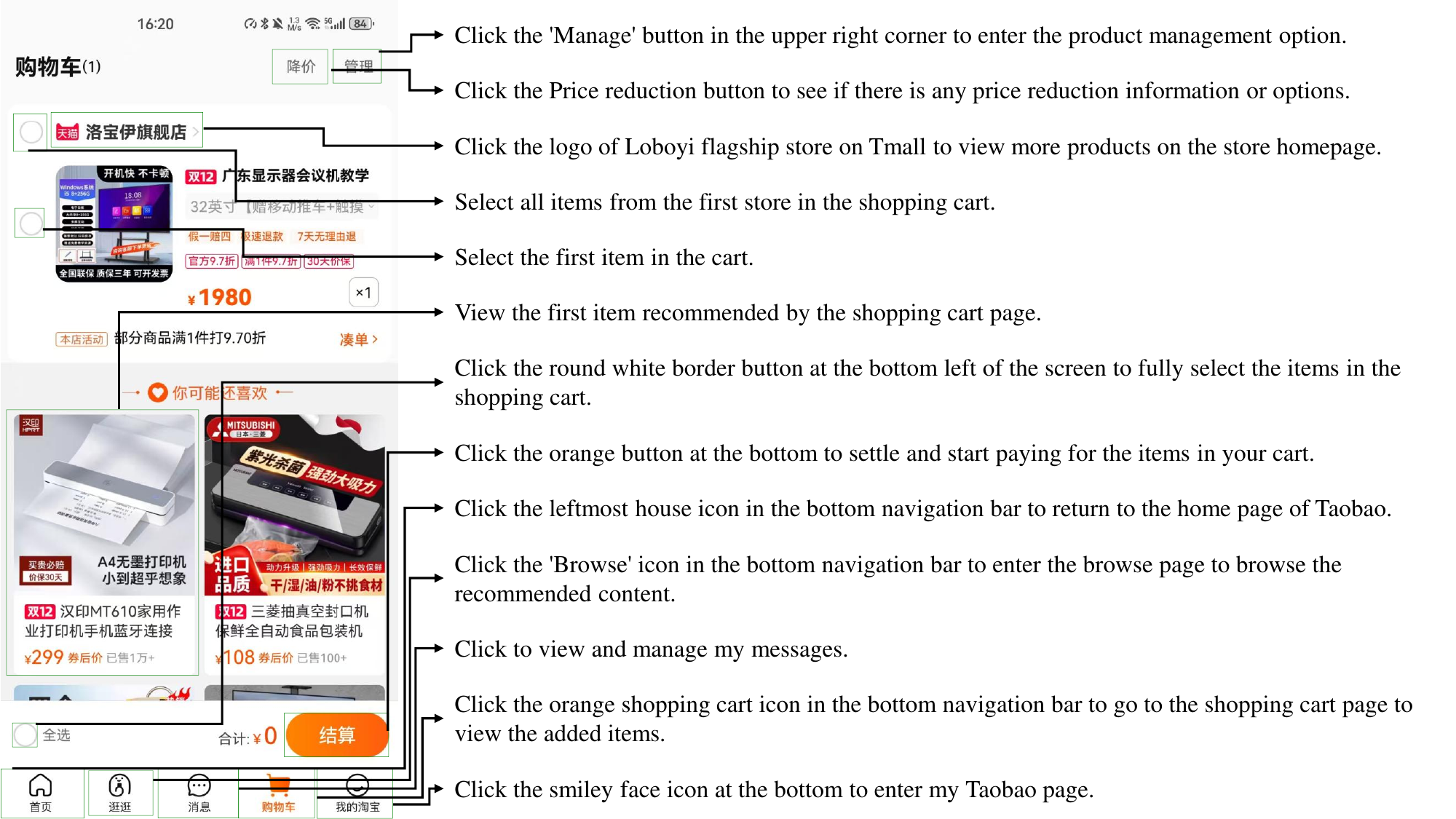}
    \caption{An example of Android version.}
    \label{fig:example_1}
\end{figure}

\begin{figure}[!h]
    \centering
    \includegraphics[width=.99\linewidth]{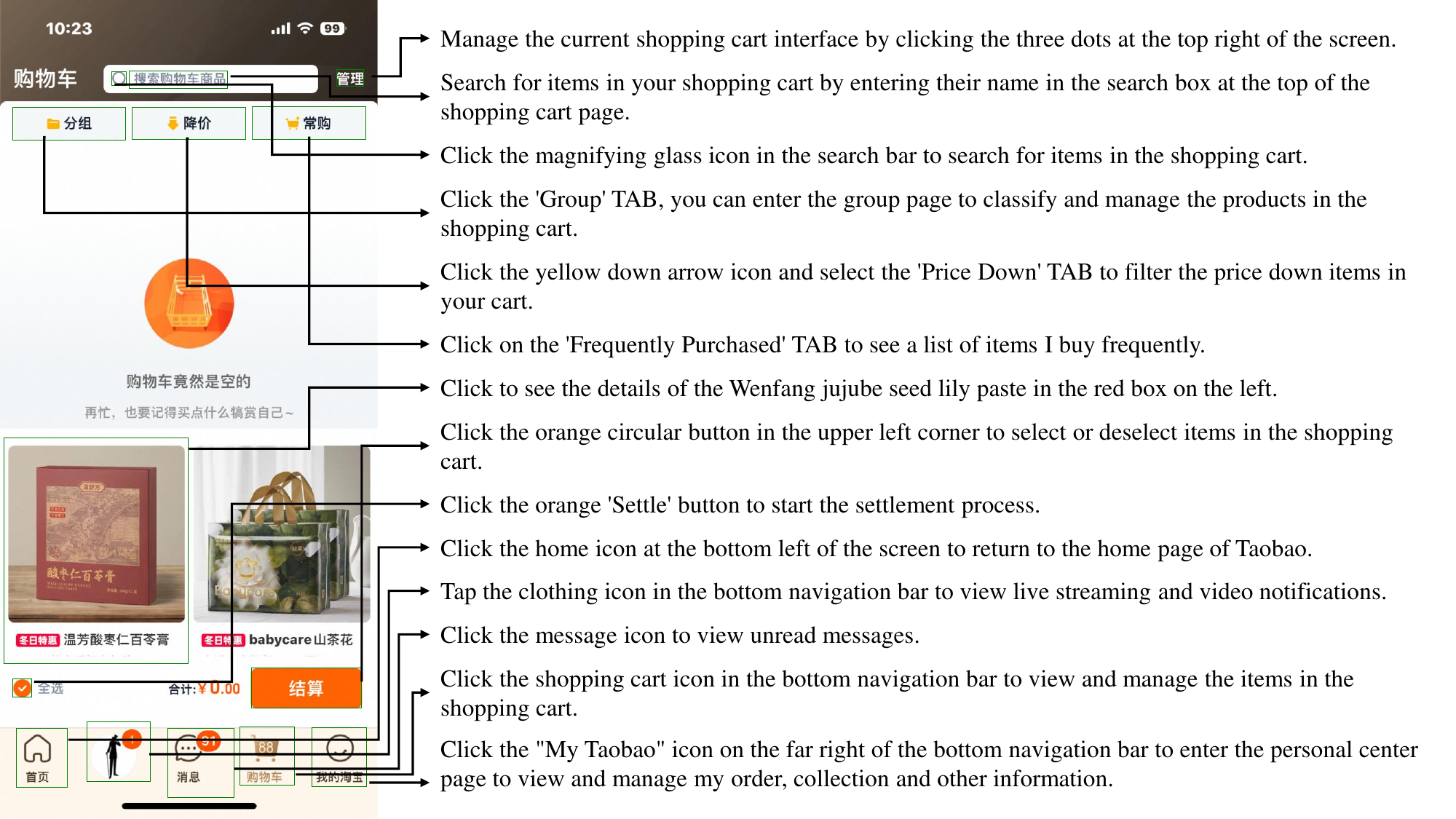}
    \caption{An example of iOS version.}
    \label{fig:example_3}
\end{figure}

\begin{figure}[!ht]
    \centering
    \includegraphics[width=.9\linewidth]{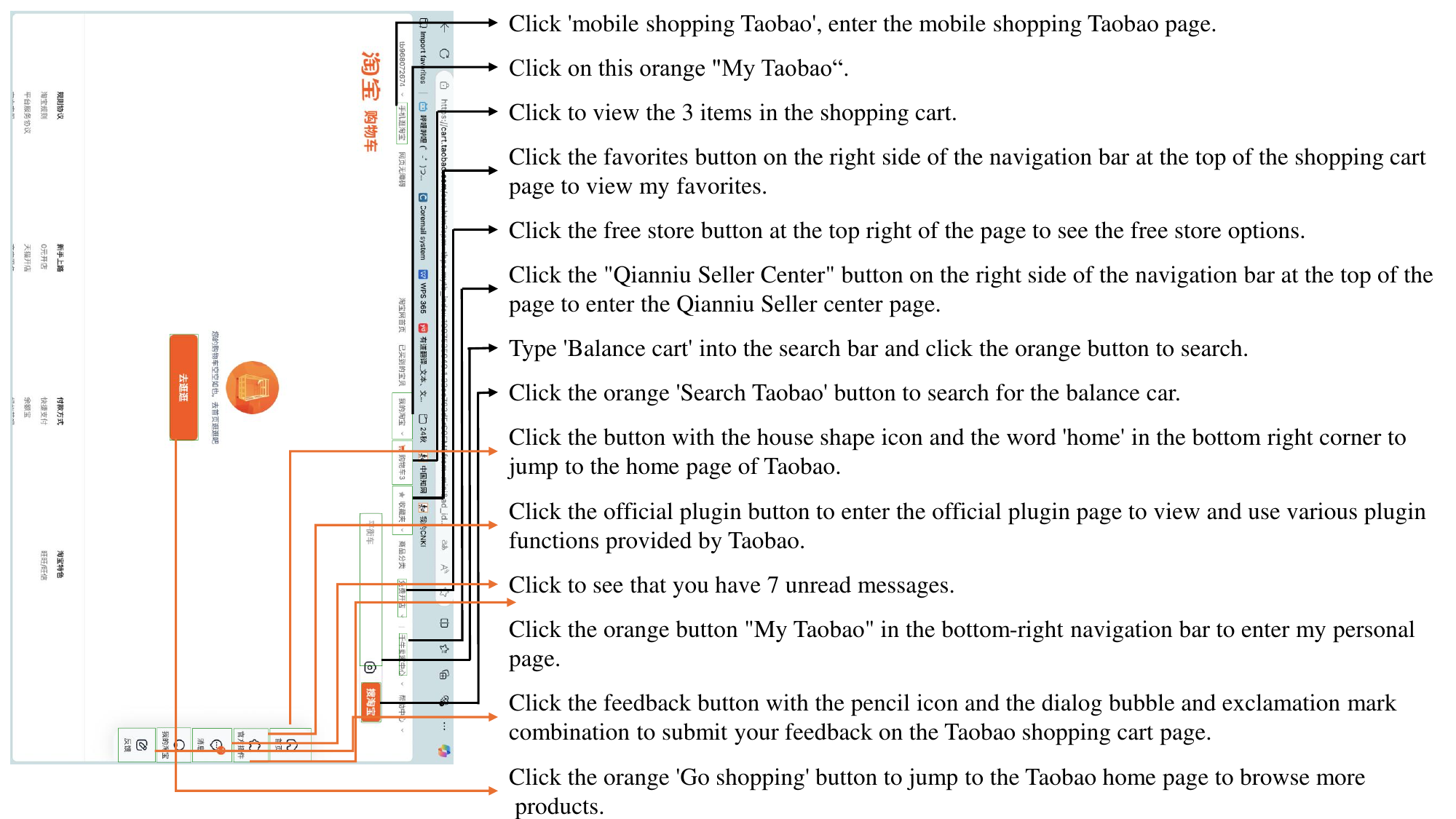}
    \caption{An example of the Web version.}
    \label{fig:example_4}
\end{figure}

Here are data examples (Figure \ref{fig:example_1}, \ref{fig:example_3}, and \ref{fig:example_4}) for three platforms: Android new, iOS, and Web. In each example, there are several bounding boxes and the corresponding instructions. It is worth noting that the different colors are only for convenience in viewing more details and do not have different semantics.

\vspace{-8pt}


\subsection{Application Names and Page Titles}
\label{appendix1_name_title}

Table \ref{tab:app_name_desc} and \ref{appendix:page_title_desc} respectively list each category's application names as well as page titles (comprised of both fundamental pages and domain-specific pages).

\begin{table*}[!ht]
\centering
    \begin{adjustbox}{max width=0.9 \textwidth}
    
    \begin{tabular}{l | cccc}
    \hline
    \textbf{App Categories} & \textbf{APP Names} \\
    \hline
    Shopping & \textit{Taobao}, \textit{Jingdong}, \textit{Weipinhui}, \textit{TMall}, \textit{Pinduoduo}, \textit{Dewu}, \textit{Yitao}, \textit{Alibaba1688}\\
    Video Streaming & \textit{Aiqiyi}, \textit{Tencent Video,}, \textit{Youku Video}, \textit{bilibili}, \textit{Mangou}, \textit{Xigua}, \textit{Tudou}, \textit{Souhu} \\
    Social Networking & \textit{Douyin}, \textit{Xiaohongshu}, \textit{Kuaishou}, \textit{Douyinjisu}, \textit{Douyinhuoshan}, \textit{Weibo}, \textit{Jinritoutiao} \\
    Travel & \textit{Qunaer}, \textit{Xiecheng}, \textit{12306}, \textit{Tongcheng}, \textit{Feizhu}, \textit{Zhixing}, \textit{Tuniu}\\
    Lifestyle & \textit{eleme}, \textit{meituan}, \textit{dingdong}, \textit{Hema}, \textit{Didi}, \textit{Dazhong}\\
    Maps &  \textit{Gaode Maps}, \textit{Baidu Maps}, \textit{Tencent Maps}, \textit{Beidou} \\
    Music & \textit{Netease}, \textit{QQ Music}, \textit{Kugou Music}, \textit{Qishui Music}, \textit{Migu}, \textit{Kuwo}, \textit{Bodian}, \textit{Quanmin}\\ 
    Communication &  \textit{QQ}, \textit{Feishu}, \textit{DingTalk} \\
    Finance &  \textit{Bank of China}, \textit{Bank of Construction}, \textit{Alipay} \\
    Email & \textit{QQ Mail}, \textit{Netease Mail}, \textit{189 Mail} \\
    Reading & \textit{Kindle}, \textit{Wechat Reading}, \textit{Fanqie}, \textit{Douban}, \textit{Qimao}, \textit{Netease Reading} \\
    Education & \textit{Wanciwang}, \textit{Xindongfang}, \textit{Momo}, \textit{Hujiang}, \textit{Baicizhan} \\
    Camera & \textit{Huangyou}, \textit{Qingyan}, \textit{Meitu}, \textit{B612}, \textit{Xingtu} \\
    Fitness & \textit{Keep}, \textit{MeiriYoga}, \textit{Yinghan} \\
    Utility Tools & \textit{Fanqie}, \textit{Ticktick} \\
    \hline
    \end{tabular}
    \end{adjustbox}
\caption{List of all Apps and their corresponding names in \texttt{TransBench}}
\label{tab:app_name_desc}
\end{table*}

\begin{table*}[!ht]
\centering
    \begin{adjustbox}{max width=0.9 \textwidth}
        \begin{tabular}{l|ll}
        \hline
        \multicolumn{1}{l|}{} & \multicolumn{2}{c}{\textbf{Page Titles}}  \\ \cline{2-3} 
        \multicolumn{1}{l|}{\multirow{-2}{*}{\textbf{App Categories}}} & \multicolumn{1}{l}{Fundamental pages} & \multicolumn{1}{l}{Domain-specific pages}      \\ 
        \hline
        Shopping & \textit{Home}, \textit{Me}, \textit{Message} & \textit{Cart}, \textit{Orders} \\
        Video Streaming & \textit{Home}, \textit{Me}, \textit{Search} & \textit{Video}, \textit{Full Screen}, \textit{History}, \textit{Advertisement} \\
        Social Networking & \textit{Home}, \textit{Me}, \textit{Search} & \textit{Full Screen}, \textit{Comments} \\
        Travel & \textit{Home}, \textit{Me} & \textit{Orders}, \textit{Booking}, \textit{Search}, \textit{Flights} \\
        Lifestyle & \textit{Home}, \textit{Me} & \textit{Cart}, \textit{Recommendations}, \textit{Search}, \textit{Details}, \textit{Orders} \\
       \textbf{ }Maps & \textit{Home} & \textit{Details} \\
        Music & \textit{Home}, \textit{Me}, \textit{Search} & \textit{Video}, \textit{Comments}, \textit{Favorites} \\
        Communication & \textit{Home}, \textit{Me} & \textit{Profile}, \textit{Settings}, \textit{Contacts}, \textit{Moments}, \textit{More} \\
        Finance & \textit{Home}, \textit{Me}, \textit{Search} & \textit{Customer Service} \\
        Email & \textit{Home} & \textit{Inbox}, \textit{Emails}, \textit{Compose} \\
        Reading & \textit{Home}, \textit{Me}, \textit{Search} & \textit{Details} \\
        Education & \textit{Home}, \textit{Search} & \textit{Details}  \\
        Camera & \textit{Home} & \textit{Photo}, \textit{Edit} \\
        Fitness & \textit{Home}, \textit{Me}, \textit{Search} & \textit{Start Exercise} \\
        Utility Tools & \textit{Home}, \textit{Me} & \textit{Data Statistics}, \textit{Add to-do Items} \\
        \hline
        \end{tabular}
    \end{adjustbox}
\caption{List of all App categories and their corresponding page titles in \texttt{TransBench}}
\label{appendix:page_title_desc}
\end{table*}

\begin{figure}[!h]
    \centering
    \includegraphics[width=0.98\linewidth]{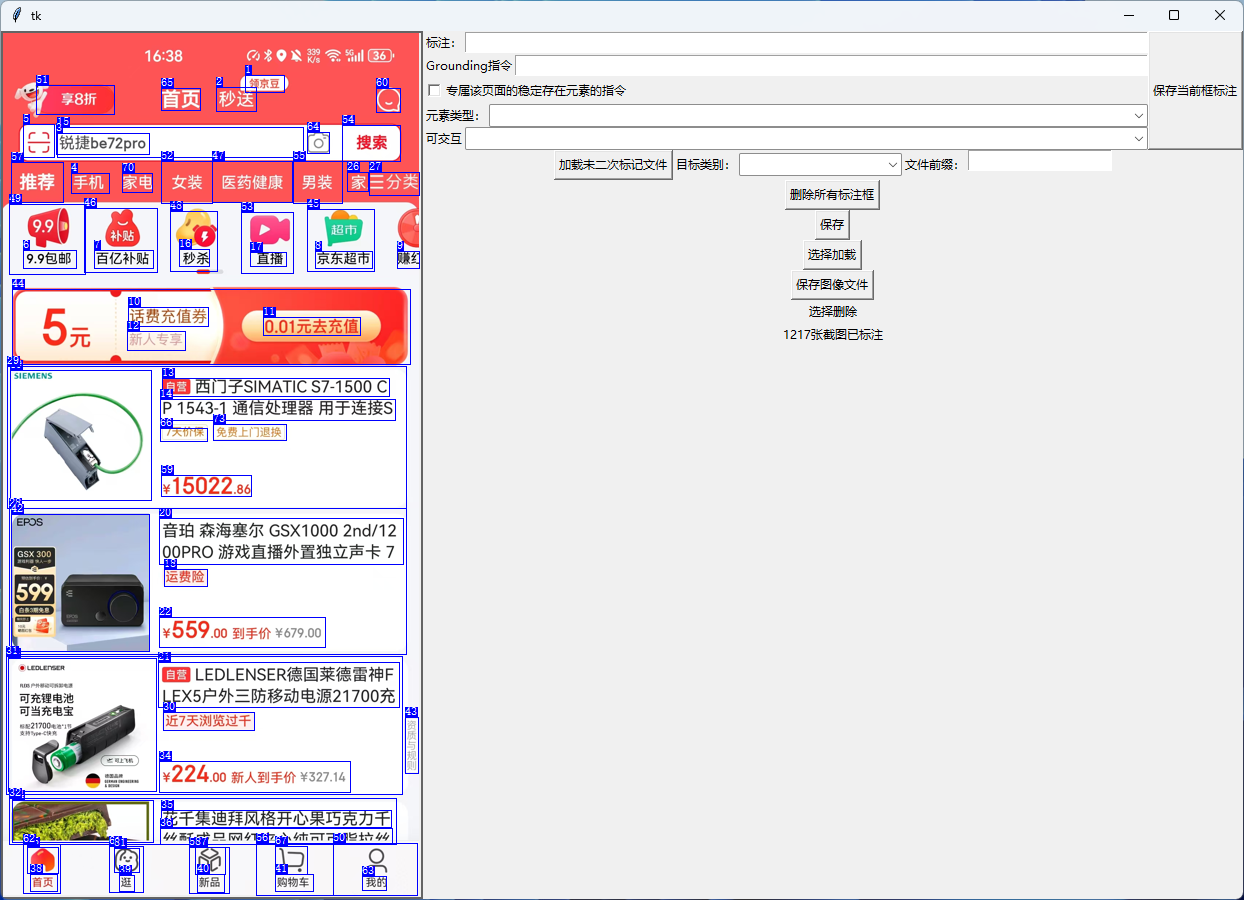}
    \caption{Enter Caption}
    \label{fig:guilabeller}
\end{figure}

\subsection{Other details}
\label{appendix1_tool_detail}
\paragraph{Tools details.} To quickly inspect the generated data, we develop a manual inspection tool in Figure \ref{fig:guilabeller} called GUILabeller, which uses Python and can run across different platforms. We will open-source this tool on GitHub.

\paragraph{Data collection pipeline details.} To accelerate the generation speed, as well as due to resource constraints, we utilized a combination of online Qwen series models (including qwen-VL-plus, qwen-VL-max, and qwen-plus) and locally run Qwen2VL-72b model and Qwen2-72b model. Specifically, the locally run models are running on an NVIDIA A800 server cluster using int4 format quantization.

\section{Experiments}
\label{appendix:exp}

\subsection{Evaluation Details}
\label{appendix2_prompt_details}
To ensure optimal performance for each model, all hyperparameters (e.g., temperature) are set consistently with their publicly released versions. We adapt our evaluation framework to each model and provide the evaluation scripts in our repository. Since different models utilize distinct prompt structures during training, we strictly follow the prescribed prompt formats for each model to achieve their best performance. Specifically, CogAgent requires the platform type as input, and we provide the correct platform type accordingly. We will open-source our testing scripts in the GitHub repository, which facilitates the easy addition of new agents for testing. The detailed prompts are in Table \ref{appendix:exp_prompt_1}.

\begin{table}[!htbp]
\small
    \centering
    \colorbox{blue!8}{
    \begin{tabular}{@{}p{7.2cm}}
\textbf{ARIA-UI:}\\
\\
    Given a GUI image, what are the relative (0-1000) pixel point coordinates for the element corresponding to the following instruction or description: \itshape\{instruction\} \upshape\\
\\
\textbf{CogAgent:}\\
\\
Task: Click on the element most relevant to the instruction ”\itshape\{instruction\} \upshape”\\
History steps: \\
(platform: \itshape\{platform\}\upshape)\\
(Answer in Status-Action-Operation-Sensitive format.)\\
\\
\textbf{OS-Atlas:}\\
\\
In this UI screenshot, what is the position of the element corresponding to the command "\itshape\{instruction\} \upshape" (with box)?\\
\\
\textbf{Qwen2.5VL:}\\
\\
The user query: Please click the most suitable“\itshape\{instruction\} \upshape”element:\\
\\
\textbf{SeeClick:}\\
\\
In this UI screenshot, what is the position of the element corresponding to the command "\itshape\{instruction\} \upshape"(with point)?\\
\\
\textbf{UGround:}\\
\\
Your task is to help the user identify the precise coordinates (x, y) of a specific area/element/object on the screen based on a description.\\
- Your response should aim to point to the center or a representative point within the described area/element/object as accurately as possible.\\
- If the description is unclear or ambiguous, infer the most relevant area or element based on its likely context or purpose.\\
- Your answer should be a single string (x, y) corresponding to the point of the interest.\\
Description: \itshape\{instruction\} \upshape\\
Answer:\\\end{tabular}
    }
    \caption{The model evaluation prompts used on LLMs. Qwen2.5VL has long system message, which follows the message in Qwen2.5VL repository.}
    \label{appendix:exp_prompt_1}
\end{table}

\subsection{Training Set Division}
\label{appendix2_trainingset_division}
Our dataset is annotated with five key dimensions: app names, app categories, page titles, app versions, and platform types, enabling multi-dimensional partitioning. As illustrated in the figure below, models fine-tuned on specific partitions are exclusively evaluated on their corresponding test sets to prevent data leakage.
\paragraph{Android-Low Partition:} From the dataset, 5,696 low-version Android samples are selected as candidates. We randomly chose 5,000 for training, with the remaining 696 and an additional 16,542 samples used for testing.

\paragraph{iOS Partition:} A total of 6,046 iOS samples are filtered as candidates. We randomly select 5,000 for training, leaving 1,046 and an additional 16,192 samples for testing.

\paragraph{Web Partition:} From 4,191 Web platform samples, 4,000 are randomly chosen for training, with the remaining 191 and an additional 18,047 samples used for testing.

\paragraph{Normal Partition: }We randomly select 5,000 samples from the entire dataset for training, using the remaining data for testing.

\paragraph{App Partition:} To evaluate cross-app transferability, we first select the top 7 app categories with the most data. From these, 40\% of the apps are reserved for testing, resulting in 6,247 candidate samples. We then randomly select 5,000 for training, with the remaining 1,247 and an additional 15,991 samples used for testing.

\subsection{Finetuning}
\label{appendix2_finetuning}

We follow the Lora fine-tuning parameters officially provided by Aria and perform Lora fine-tuning. Specifically, we trained Lora with r=8, alpha=32, dropout=0.05, and target\_modules as "fc1", "fc2", "q\_proj", "k\_proj", "v\_proj", "linear", "o\_proj", "up\_proj", "down\_proj", "out\_proj", "gate\_proj", "lm\_head". The learning rate was set to 5e-5, the batch size was set to 16, and a total of 2 epochs were trained.

\begin{figure}[!htbp]
    \centering
    \includegraphics[width=1\linewidth]{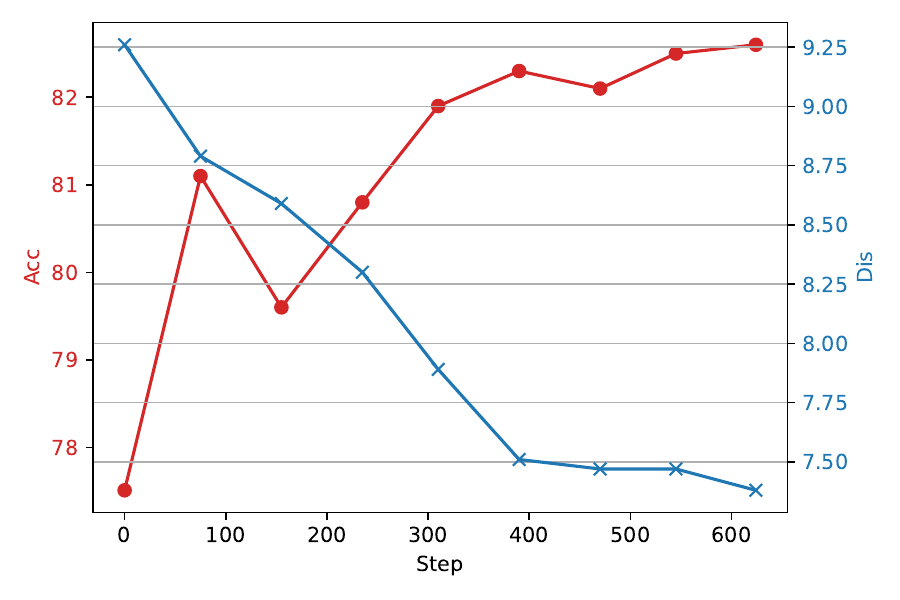}
    \caption{Variation of Average Accuracy and Average Distance on the Test Partition with Respect to Fine-Tuning Steps on the Normal Training Partition.}

    \label{fig:valid_epoch}
\end{figure}

To validate the reasonableness of selecting two epochs, we documented the accuracy and distance variations in the validation set during fine-tuning Aria-ui on the normal partition, as illustrated in Figure \ref{fig:valid_epoch}. The results indicate that two epochs essentially achieve the model's optimal performance, with minimal gains from further training, which could potentially lead to overfitting.

All fine-tuning and experiments were conducted for 200 GPU hours on the NVIDIA A800 cluster. Aria-ui has 25B parameters. We set seed as 42 for shuffle, using default config of transformers Trainer. We also plan to open-source our training scripts. 

\subsection{Error Analysis}
\label{appendix2_error_analysis}
\begin{figure}[!htbp]
    \centering
    \includegraphics[width=1\linewidth]{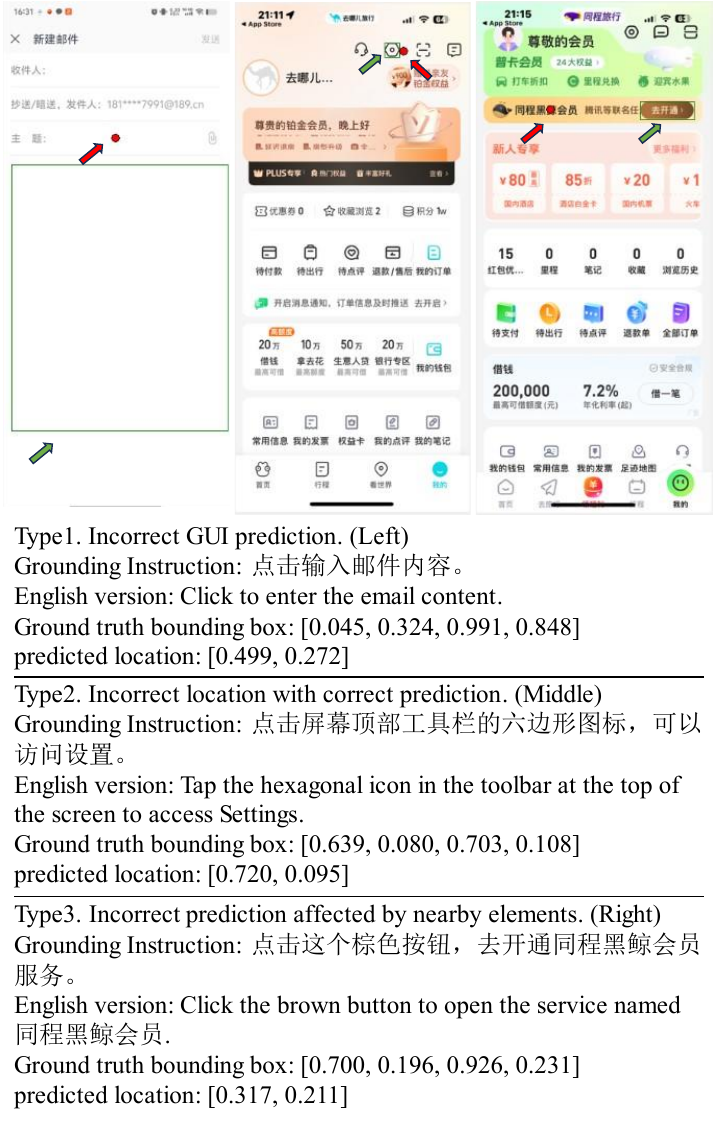}
    \caption{Three examples of error cases.}
    \label{app:error}
\end{figure}

The failure cases can be summarized into three categories, illustrated in Figure \ref{app:error}:

\paragraph{Incorrect GUI prediction.} Models do not understand grounding instructions and predict wrong click positions.
\paragraph{Incorrect location with correct prediction.} Despite the ability to understand grounding instructions, models generate predicted positions that fall near the bounding boxes.
\paragraph{Incorrect prediction affected by nearby elements.} While models predict the target as a whole, instructions focus on only a specific part of it.

\end{document}